\documentclass[fleqn,usenatbib]{mnras}

\usepackage{mathptmx}

\usepackage[T1]{fontenc}

\DeclareRobustCommand{\VAN}[3]{#2}
\let\VANthebibliography\thebibliography
\def\thebibliography{\DeclareRobustCommand{\VAN}[3]{##3}\VANthebibliography}

\usepackage{graphicx}	
\usepackage{amsmath}	
\usepackage{txfonts}
\usepackage{subeqnarray}
\usepackage{caption}
\usepackage{subcaption}
\usepackage{url}
\usepackage{threeparttable}
\usepackage{multirow}

\newcommand{\kms}{km.s$^{-1}$}
\newcommand{\ms}{m.s$^{-1}$}

\newcommand{\kp}{K$_{\rm{p}}$}
\newcommand{\vsys}{V$_{\rm{sys}}$}

\newcommand{\hdb}{HD~189733\,b}
\newcommand{\mdw}{M\,dwarf}
\newcommand{\GLA}{Gl~15~A}

\defcitealias{Maciejewski2014}{Ma14}
\defcitealias{rosenthal2021}{Ro21}
\defcitealias{cutri2003}{Cu03}
\defcitealias{stassun2017}{St17}
\defcitealias{addison2019}{Ad19}
\defcitealias{bourrier2018}{Bo18}
\defcitealias{baluev2015}{Ba15}
\defcitealias{trifonov2021}{Tr21}
\defcitealias{claret2011}{Cl11}
\defcitealias{fouque2018}{Fo18}
\defcitealias{turner2016}{Tu16}
\defcitealias{cristofari2022}{Cr22}

\title[ATMOSPHERIX: HRS with SPIRou]{ATMOSPHERIX: I- An open source high resolution transmission
spectroscopy pipeline for exoplanets atmospheres with SPIRou}

\author[Klein, B. et al.]{
Baptiste Klein,$^{1,2}$\thanks{E-mail: baptiste.klein@physics.ox.ac.uk}
Florian Debras,$^{1}$
Jean-François Donati, $^{1}$
Thea Hood,$^{1}$
Claire Moutou,$^{1}$
Andres Carmona, $^{3}$\newauthor
Merwan Ould-elkhim,$^{1}$
Bruno Bézard, $^{4}$
Benjamin Charnay,$^{4}$
Pascal Fouqué, $^{1}$
Adrien Masson,$^{4}$\newauthor
Sandrine Vinatier,$^{4}$
Clément Baruteau, $^{1}$
Isabelle Boisse,$^{5}$
Xavier Bonfils,$^{3}$
Andrea Chiavassa,$^{6}$
Xavier Delfosse,$^{3}$\newauthor
William Dethier,$^{3}$
Guillaume Hebrard,$^{7,8}$
Flavien Kiefer,$^{7}$
J\'er\'emy Leconte,$^{9}$
Eder Martioli,$^{7,10}$\newauthor
Vivien Parmentier, $^{2,6}$
Pascal Petit,$^{1}$
William Pluriel,$^{11}$
Franck Selsis,$^{9}$
Lucas Teinturier,$^{4,13}$
Pascal Tremblin,$^{12}$\newauthor
Martin Turbet, $^{13,9}$
Olivia Venot,$^{14}$
and Aurélien Wyttenbach$^{3}$
\\
$^{1}$IRAP, Université de Toulouse, CNRS, UPS, Toulouse, France\\
$^{2}$Department of Physics, University of Oxford, OX13RH, Oxford, UK\\
$^{3}$Université Grenoble Alpes, CNRS, IPAG, 38000 Grenoble, France\\
$^{4}$LESIA, Observatoire de Paris, Université PSL, Sorbonne Université, Université Paris Cité, CNRS, 5 place Jules Janssen, 92195 Meudon, France\\
$^{5}$Aix Marseille Universite, CNRS, Laboratoire d’Astrophysique de Marseille UMR 7326, 13388, Marseille, France\\
$^{6}$Université Côte d'Azur, Observatoire de la Côte d'Azur, CNRS, Lagrange, CS 34229, Nice, France \\
$^{7}$Institut d'astrophysique de Paris, UMR7095 CNRS, Universit\'e Pierre \& Marie Curie, 98bis boulevard Arago, 75014 Paris, France\\
$^{8}$Observatoire de Haute-Provence, CNRS, Universit\'e d'Aix-Marseille, 04870 Saint-Michel-l'Observatoire, France\\
$^{9}$Laboratoire d'astrophysique de Bordeaux, Univ. Bordeaux, CNRS, B18N, all\'ee Geoffroy Saint-Hilaire, 33615 Pessac, France \\
$^{10}$Laborat\'{o}rio Nacional de Astrof\'{i}sica, Rua EstadosUnidos 154, 37504-364, Itajub\'{a} - MG, Brazil \\
$^{11}$Observatoire astronomique de l'Universit\'e de Gen\`eve, chemin Pegasi 51, 1290 Versoix, Switzerland \\
$^{12}$Universite Paris-Saclay, UVSQ, CNRS, CEA, Maison de la Simulation, 91191, Gif-sur-Yvette, France\\
$^{13}$Laboratoire de Météorologie Dynamique/IPSL, CNRS, Sorbonne Université, Ecole Normale Supérieure, Université PSL, Ecole Polytechnique,\\
Institut Polytechnique de Paris, 75005 Paris, France \\
$^{14}$Université de Paris Cité and Univ Paris Est Creteil, CNRS, LISA, F-75013 Paris, France\\
}

\date{Accepted XXX. Received YYY; in original form ZZZ}

\pubyear{2022}

\begin{document}
\label{firstpage}
\pagerange{\pageref{firstpage}--\pageref{lastpage}}
\maketitle

\begin{abstract} Atmospheric characterisation of exoplanets from the ground is an actively growing field of research. In this context we have created the ATMOSPHERIX consortium: a research project aimed at characterizing exoplanets atmospheres using ground-based high resolution spectroscopy. This paper presents the publicly-available data analysis pipeline and demonstrates the robustness of the recovered planetary parameters from synthetic data.  Simulating planetary transits using synthetic transmission spectra of a hot Jupiter that were injected into real SPIRou observations of the non-transiting system \GLA, we show that our pipeline is successful at recovering the planetary signal and input atmospheric parameters. We also introduce a deep learning algorithm to optimise data reduction which proves to be a reliable, alternative tool to the commonly used principal component analysis. We estimate the level of uncertainties and possible biases when retrieving parameters such as temperature and composition and hence the level of confidence in the case of retrieval from real data. Finally, we apply our pipeline onto two real transits of HD~189733 b observed with SPIRou and obtain similar results than in the literature. In summary, we have developed a publicly available and robust pipeline for the forthcoming studies of the targets to be observed in the framework of the ATMOSPHERIX consortium, which can easily be adapted to other high resolution instruments than SPIRou (e.g. VLT-CRIRES, MAROON-X, ELT-ANDES). 
\end{abstract}

\begin{keywords}
exoplanets -- planets and satellites: atmospheres -- planets and satellites: gaseous planets -- techniques: spectroscopic -- methods: data analysis
\end{keywords}



\section{Introduction}\label{sec:intro}

More than 5\,000 exoplanets were discovered in the last decade, paving the way for statistical exploration of the orbital and physical properties of planetary systems \citep[e.g.,][]{Udri2007,Fulton2017,Debras2021}. One of the main objective for the next decade is now to understand thoroughly the physical nature of individual planets. This necessarily requires an in-depth study of their atmosphere in order to lift degeneracies between seemingly identical planets in terms of mass and radius \citep[e.g.,][]{Valencia2013}. JWST and Ariel \citep{Tinetti2021} space missions will play a key role in that venture by providing high quality observations for a large number of exoplanets. However, space-based observations of planet atmospheres have limits which are best overcome from the ground using high-resolution spectroscopy (HRS) with numerous large telescopes. Through cross-correlation high resolution spectroscopy, relying on the statistical comparison between an absorption or emission spectrum of the planet atmosphere and theoretical models, one can extract the planetary signal which is typically 10 to 100 times weaker than the noise. Since the first successful characterisation of an exoplanet atmosphere with high-resolution spectroscopy, a decade ago by \citet{Snellen2010}, this technique has been substantially refined \citep[e.g.,][]{brogi2012,dekok13,birkby2013,Brogi2016,alonso-floriano2019,Brogi2019,Giacobbe2021,guilluy2022}, and has acquired the necessary maturity to become a reliable complement to forthcoming space-based missions \citep{Brogi2017,Brogi2019,kasper2023}.


SPIRou \citep{Donati2020}, a high-resolution near-infrared (nIR) spectropolarimeter mounted at the Canada-France-Hawaii Telescope, is perfectly suited for this task. Thanks to its broad continuous wavelength coverage of the near-infrared, from 0.9 to 2.5\,$\mu$m, SPIRou has the ability to resolve a high number of molecular lines in the emission or transmission spectra of planetary atmospheres. More specifically, the observations are divided into 50 overlapping diffraction orders spanning the $Y$,$J$,$H$ and $K$ bands at a resolving power of $\sim$70\,000 ($\sim 2.28$ km.s$^{-1}$ velocity bin). SPIRou has already been successfully used to detect water and carbon monoxide by performing emission spectroscopy of $\tau$ Boo b \citep{Pelletier2021} and transmission spectroscopy during two transits of HD 189733 b \citep{Boucher2021} as well as to detect Helium on several targets \citep[][Masson et al. in prep.]{allart2023}. In most cases, the absorption lines of the planet atmosphere were found to be Doppler shifted compared to theoretical predictions which remains to be understood in the light of atmospheric circulation \citep[see e.g.][]{Flowers2019}.


With the aim of optimising the capabilities of SPIRou for the characterisation of the atmosphere of exoplanets, we have gathered a large, French-led community of observers and theoreticians, specializing in exoplanet atmospheres and stellar observations and simulations, under the ATMOSPHERIX program. This program aims at observing a wide range of exoplanets over several years in order to (i)~constrain the composition of their atmosphere, (ii)~probe the pressure-temperature (PT) profile and the amplitude of atmospheric winds through Doppler spectroscopy, and (iii)~characterise the extended atmosphere through the He\,I metastable triplet at 1083\,nm. Additionally, long-term repeated observations of a sample of planets will allow us to better understand variability in exoplanet atmospheres \citep[e.g.,][]{Armstrong2016,Komacek2020,Cho2021}.

This paper introduces a series of studies of the atmosphere of transiting planets observed with nIR high-resolution spectrographs as part of the ATMOSPHERIX program. In this study, we present our publicly-available  code\footnote{\url{https://github.com/baptklein/ATMOSPHERIX_DATA_RED}} to extract a planet transmission spectrum from a time-series of nIR high-resolution spectra. The extraction pipeline is applied to  (i) sequences of synthetic transmission spectra of a hot Jupiter that mimics the properties of \hdb\, and that was injected into SPIRou observations of the bright quiet \mdw\ \GLA\ and (ii) on the two same transits of HD 189733 b as \citet{Boucher2021}. The \GLA\ input data sets are described in Section~\ref{sec:Observations}. Section~\ref{sec:Methods} details the algorithm to extract the planet atmosphere signal from the observed sequence of spectra and infer the planet atmosphere parameters in a statistically-robust way. We then present our retrieval methods on synthetic data in Section~\ref{sec:Results} and their application on real data in Section \ref{sec:HD189}.  We discuss our results and their implications for real targets in Section \ref{sec:disc} and conclude in Section \ref{sec:conclusion}. A companion paper (Debras et al., submitted to MNRAS) studies the biases and degeneracies in the planet atmosphere parameters retrieved with the pipeline presented here.

\section{Observations and planet injection}\label{sec:Observations}

We simulate nIR spectroscopic observations of a planetary transit by injecting a synthetic planet atmosphere spectrum into a sequence of spectra of the bright M dwarf \GLA, collected with SPIRou in October 2020 (see Table\,\ref{tab:parameters}). We first describe the stellar data before detailing how we injected the planet. 

\subsection{Input stellar spectra}

\GLA\ has been intensively monitored with SPIRou over the last 3 years and does not have any known transiting planet, making it an ideal target to benchmark our data analysis code. Our input observations consist in a series of 192 consecutive spectra collected with SPIRou on October 8, 2020, spanning a total of 5 hours. We divided these 192 spectra into two series of 96 spectra (the odd and even file numbers, respectively) in order to ensure the robustness of our pipeline on two sets of data. The peak signal-to-noise ratio (SNR) per 2.28 \kms\ velocity bin ranges from 250 to 320 (median of 291), and the airmass from 1.1 and 1.3 (see panels 2 and 4 of Figure~\ref{fig:transit_info}). Note that the binary companion, Gl 15 B, is a M\,3.5 dwarf located at 146\,au from \GLA\ \citep{reid1995}. The velocimetric effect of this binary system is neglected in the present analysis given the low acceleration that B induces on A's RV \citep[about 2\ms per year;][]{howard2014}. We also neglect the RV effect of the two recently-detected planets around \GLA\ \citep{howard2014,pinamonti2018}, inducing respective signatures of 1.68 and 2.5\,\ms\ modulated with orbital periods of 11.44 and 7600\,d, respectively.

For this paper, the SPIRou observations were reduced using the version 0.6.132 of \texttt{APERO}, the official data reduction software (DRS) of the instrument \citet{cook2022}. In short, the pipeline applies the optimal extraction method of \citet{Horne1986} to extract each individual exposure from the H4RG detector \citep{Artigau2018}. The wavelength solution is obtained by combining calibration exposures of a UNe hollow-cathod lamp and a thermally-stabilised Fabry-Periot \'etalon \citep{bauer2015,hobson2021}. \texttt{\texttt{APERO}} performs a correction of the telluric contamination using a method, summarised in \citet[][see Section~8]{cook2022}, which will be presented in aforthcoming paper (Artigau et al. in prep.). This technique applies \texttt{TAPAS} \citep{bertaux2014} to pre-clean telluric absorption and the low level residuals are removed in using a data set of spectrum of hot stars observed in different atmospheric conditions to build a residual  models in function of few parameters (optical depths of H$_2$O and of dry components). Note that the deepest telluric lines (relative absorption larger than 90\%) are masked out by the pipeline as the low amount of transmitted flux will most likely result in an inaccurate telluric correction. Our input sequence of spectra contains the blaze- and telluric-corrected spectra.

\begin{table*}
\caption{Physical parameters for \GLA, HD189733 b and for the simulated hot Jupiter used in the study. When taken from the literature, the reference for each parameter is indicated in the right-hand column$^{\dagger}$. Note that references cited for planet parameters refer to HD189733 b.}
\label{tab:parameters}
\centering
\begin{threeparttable}
\begin{tabular}{cccc}
  \hline
  \textbf{Stellar parameters} & \multicolumn{3}{c|}{\textbf{\GLA}}  \\
   & \multicolumn{2}{c|}{Value} & Reference \\ 
  \hline
  Mass ($M_\odot$) & \multicolumn{2}{c|}{0.398 $\pm$ 0.004} & \citetalias{cristofari2022} \\
  Radius ($R_\odot$)& \multicolumn{2}{c|}{0.388 $\pm$ 0.013} & \citetalias{cristofari2022} \\
  Effective temperature (K) & \multicolumn{2}{c|}{3603 $\pm$ 60} &  \citetalias{cristofari2022} \\
  $H$ magnitude  & \multicolumn{2}{c|}{4.476 $\pm$ 0.2} & \citetalias{cutri2003} \\
  Systemic velocity [\kms] & \multicolumn{2}{c|}{11.73 $\pm$ 0.1} &  \citetalias{fouque2018} \\
  Limb Darkening (Quadratic) & \multicolumn{2}{c|}{0.0156, 0.313} &  \citetalias{claret2011} \\
   \hline
 \textbf{Planet parameters} &  & & \\
  & HD 189733 b & Synthetic planet & Reference \\ 
  \hline
  Transit depth (\%) & 2.2 $\pm$ 0.1 & 2.2& \citetalias{addison2019}  \\
  Radius ($R_J$) & 1.142 $\pm$ 0.040 & 0.57 & \citetalias{addison2019}  \\
  Mass ($M_J$) & 1.13 $\pm$ 0.05  & 0.568 &  \citetalias{addison2019}   \\
  g (m.s$^{-2}$) & 22.45 $\pm$ 1.5 & 45.29 & From planet mass and radius \\
  Orbital period (d)  & 2.218579 $\pm$ 0.000001& 2.218577& \citetalias{addison2019} \\
  Mid transit time (BJD TBD) & 2458334.990899 $\pm$ 0.0007 & 2459130.8962180 & \citetalias{addison2019} \\
  Inclination (deg) & 85.3 $\pm$ 0.2 & 90.0 & \citetalias{addison2019} \\
  Eccentricity & $\sim$0.0  & 0.0 & -- \\
  Equilibrium temperature (K) & 1203 $\pm$ 39  & 900 & \citetalias{addison2019} \& \citetalias{rosenthal2021} \\
  Orbital semi-amplitude (km.s$^{-1}$) & 151.2 $\pm$ 4.5 & 120.0 & --  \\
  Transit duration (h) & 1.84 $\pm$ 0.04  & 1.84 & \citetalias{addison2019}  \\
  \hline
\end{tabular}
     \begin{tablenotes}
     \footnotesize
     \item[$\dagger$] To gain some space in the table, we use aliases for the references. \citetalias{cristofari2022}, \citetalias{cutri2003}, \citetalias{fouque2018}, \citetalias{claret2011}, \citetalias{addison2019} and \citetalias{rosenthal2021} stand respectively for \citet{cristofari2022}, \citet{cutri2003}, \citet{fouque2018}, \citet{claret2011}, \citet{addison2019} and \citet{rosenthal2021}.
    \end{tablenotes}
    \end{threeparttable}   
\end{table*}

\subsection{Planet injection} \label{sec:planet_injection}

\begin{figure}
    \centering
    \includegraphics[width=\linewidth]{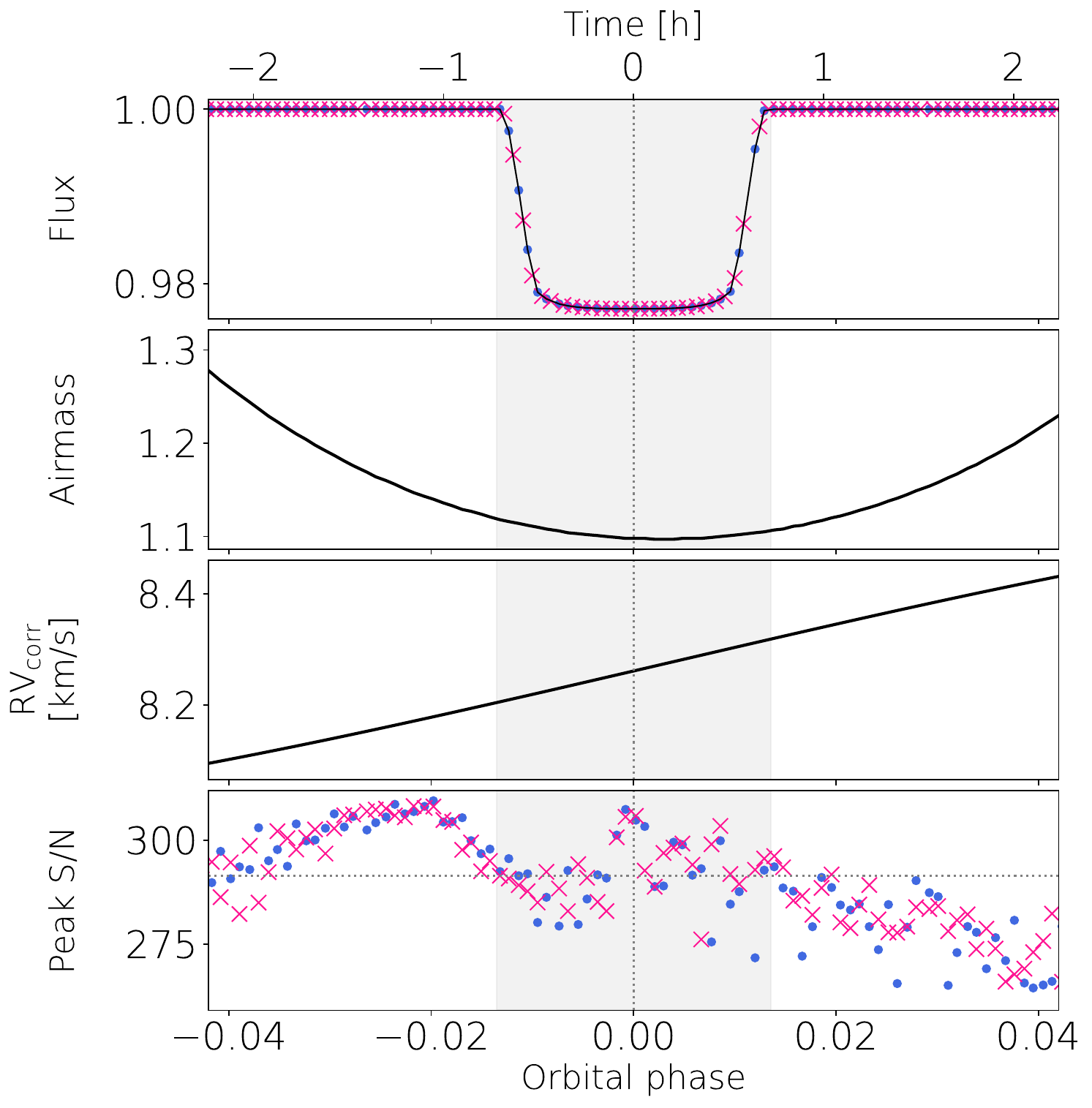}
    \caption{Continuum-normalised transit light-curve of HD 189733 b (top panel), airmass (panel~2), topocentric-to-stellar rest frame RV correction (RV$_{\rm{corr}}$, panel~3) and peak SNR per velocity bin during the two simulated transits of the \hdb\ analog (panel 4). Note that RV$_{\rm{corr}}$ contains the RV contributions of the barycentric Earth motion and of the systemic velocity of the star. On panels~1 and~4, the two different transits are respectively shown as blue dots and pink crosses. The vertical gray band (resp. vertical gray dotted line) indicates the mid-transit primary transit (resp. mid-transit time) of the simulated planet. The horizontal gray dotted line on the bottom panel indicates the average value of the peak SNR for the observed spectra.}
    \label{fig:transit_info}
\end{figure}

We then inject synthetic planet atmosphere transmission spectra on top of the APERO-provided telluric-corrected spectra of \GLA. We consider the case of a hot Jupiter \citep[based on HD 189733 b,][]{bouchy2005}. The injected planet spectra are generated using \texttt{petitRADTRANS} \citep{Molliere2019}, which gives the planet radius as a function of wavelength assuming an isothermal planet atmosphere solely containing chosen molecules at a constant volume mixing ratio. This radius is then transformed into an absorbed flux by multiplying the total flux by the ratio of planetary to stellar radius squared, called transit depth. This model is then convoluted with a Gaussian of half-width 2 SPIRou pixels \citep[i.e. $\sim$4.5\,\kms\ for a resolving power of $\sim$70\,000 in the nIR][]{Donati2020} to account for the instrumental broadening.

Since \GLA\ is significantly smaller than HD 189733, some adjustments of the injected planet parameters are needed to keep the simulations realistic. We decided to conserve 4 quantities: (i)~the transit depth, (ii)~the transit duration,  (iii)~the ratio between the stellar radius and the atmospheric scale height and (iv)~a consistent atmospheric temperature at the limbs. Our synthetic planet therefore has a lower mass, radius and velocimetric amplitude than HD189733 b, but a larger surface    gravity. Although not a physical planet (the orbital mass is not consistent with the gravitational mass), the injected planet represents a good observational analog of a hot Jupiter. The stellar properties were left untouched in our simulated data. The parameters adopted for synthetic planet are given in Table~\ref{tab:parameters}.


The planet orbit is assumed circular with a mid-transit time corresponding to the mean values of our observation times (see the two transits in Figure~\ref{fig:transit_info}). For each spectrum, we generate a transit curve $\boldsymbol{F_{\rm{C}}}$ using the python module \texttt{batman} \citep{kreidberg2015}, assuming an aligned circular planet orbit and using the H-band quadratic limb-darkening coefficients computed in \citet{claret2011} for \GLA's properties (see Table~\ref{tab:parameters}). From the resulting transiting curves, we compute transit window functions $\boldsymbol{W_{\rm{C}}}$ whose values range from 0 (for out-of-transit times) and 1 (at mid-transit time), using $\boldsymbol{W_{\rm{C}}}$\,=\,$(1-\boldsymbol{F_{\rm{C}}})/\max \left(1 - \boldsymbol{F_{\rm{C}}} \right)$.

We then build a sequence of planet transmission spectra by applying the following steps to the synthetic planet atmosphere spectrum. First, the simulated planet atmosphere spectrum is multiplied by the transit window function $\boldsymbol{W_{\rm{C}}}(t)$ at time $t$. Second, the window-weighted simulated spectrum is shifted in the Earth rest frame by correcting for the Barycentric Earth Radial Velocity $V_{\rm{BE}}$(t), the stellar systemic velocity $V_\mathrm{sys}$ and the RV signature $V_\mathrm{RV}(t)$ induced by the injected planet on the host star. We then shift this spectrum by an additional $30$ \kms\, corresponding to the planetary shift in velocity during transit plus three times the SPIRou resolution to ensure that the stellar and planetary molecular lines are separated. Note that, as \GLA\ is a M-dwarf star, it contains water and carbon monoxide in its atmosphere which complicates the planetary retrieval. This is discussed further in the companion paper. The resulting synthetic spectrum is then convolved at SPIRou's spectral resolution, and multiplied by the stellar spectrum observed at time $t$. Our input sequence of spectra is finally built by repeating the steps listed above to all the observed spectra.


The final intensity $I_\mathrm{f}$ as a function of time $t$ and wavelength $\lambda$ can thus be expressed as follows,
\begin{equation}
    I_\mathrm{f} (t,\lambda) = I_\mathrm{i} (t,\lambda) \left(1-\boldsymbol{W_{\rm{C}}}(t)\left(\dfrac{R_\mathrm{p}(\lambda'(t))}{R_\star}\right)^2 \right),
\end{equation}

\noindent
where $I_\mathrm{i}$ is the intial intensity (i.e. the \texttt{APERO} reduced SPIRou observations), $R_\star$ the stellar radius\footnote{Note that the potential wavelength-dependencies of $R_\star$ are expected to be corrected in steps (i) and (ii) of our data cleaning procedure (see Section~\ref{ssec:analysis}) and are therefore neglected in the simulations. We also assume that limb-darkening coefficients are wavelength-independent over SPIRou's spectral range.}, $R_\mathrm{p}$ the planetary radius (degraded at SPIRou resolution) which depends on wavelength because of the wavelength-dependent opacity of the planet atmosphere and $\lambda'$ the Doppler-shifted wavelength. As explained above:
\begin{gather}
    \lambda'(t) = \lambda \left( 1- \dfrac{K_\star \sin(2 \pi \phi(t))+V_\star (t)}{c_0}\right), \\
    V_\star (t) = V_{\mathrm{BE}}(t)-V_\mathrm{sys}+30 \text{km.s}^{-1}
\end{gather}

\noindent
where $K_\star$ is the orbital RV semi-amplitude of the star, $\phi$ the planet orbital phase centered on the mid transit, $V_\star(t)$ the non orbital Doppler shift and $c_0$ the speed of light.

Finally, we have also created synthetic sequences without the star, where the model is injected into a map of white noise with a variance equal to the SNR of the observations modulated by the blaze function that increases the variance at the edges of the orders. These synthetic models do not need to go through any further step of data analysis, and serve as references to identify the effects of the data analysis on the atmospheric retrieval. 

\section{Data processing}\label{sec:Methods}

\subsection{Data cleaning}\label{ssec:analysis}

\begin{figure*}
    \centering
    \includegraphics[width=\linewidth]{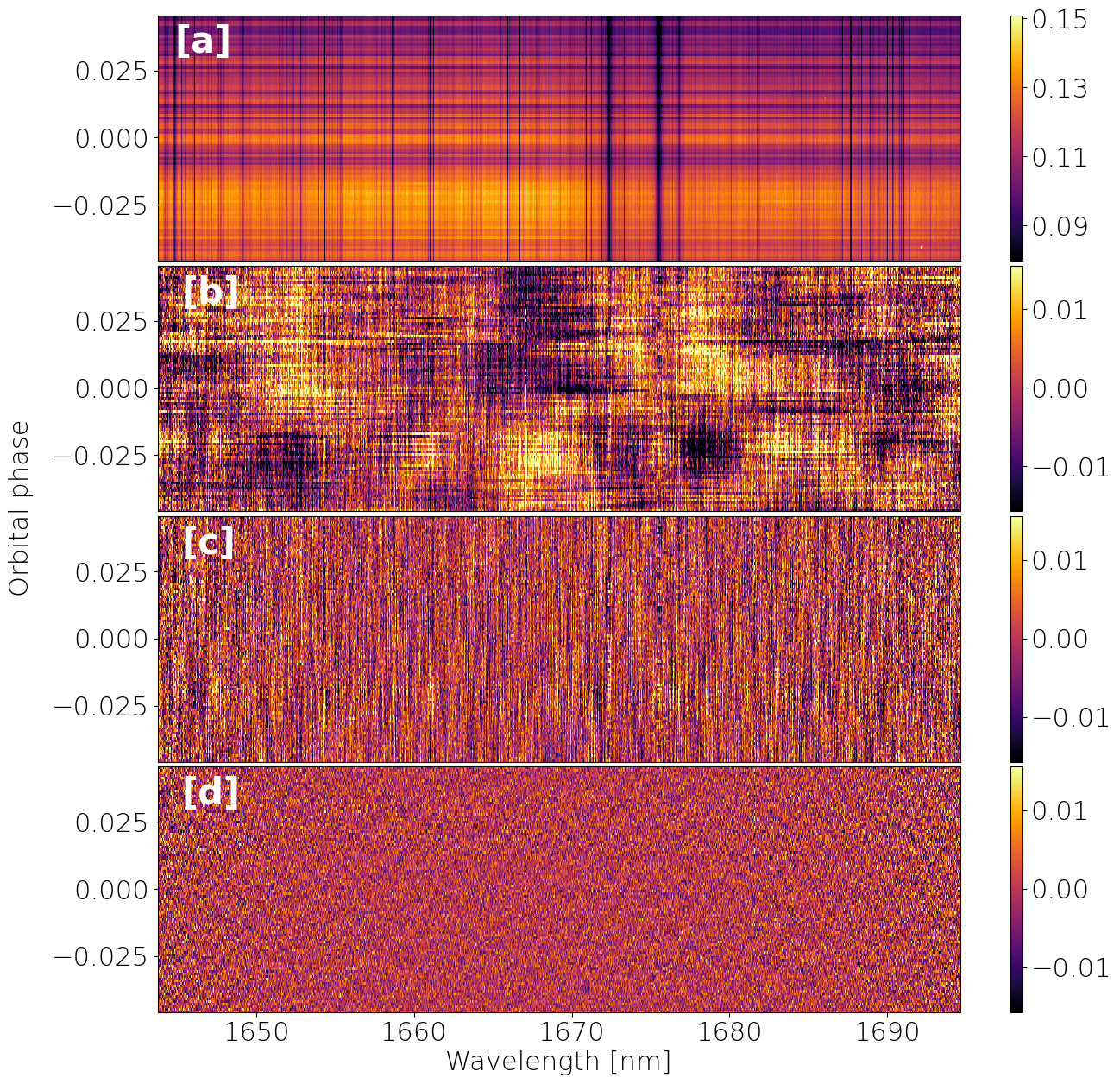}
    \caption{Time series of spectra in Order~46 (1643 to 1695\,nm) at subsequent steps of the data processing as detailed in Sections~\ref{ssec:analysis} and~\ref{ssec:pca}. From top to bottom: [a]~Blaze- and telluric-corrected APERO-provided spectra (prior to step~(i) in Section~\ref{ssec:analysis}); [b]~spectra corrected from the master-out template (step~(i) in Section~\ref{ssec:analysis}); [c]~Normalised spectra (step~(ii) in Section~\ref{ssec:analysis}); and [d]~PCA-corrected spectra (with 8 components removed; see Section~\ref{ssec:pca}). Note that time series of normalised spectra cleaned with the auto-encoder, visually similar to the bottom panel, is not shown here. }
    \label{fig:analysis}
\end{figure*}

The extraction of the planetary signal requires to correct for the residual telluric absorption lines, the stellar spectrum and additional correlated noise induced by the instrument and the observing conditions. Following \citet{Boucher2021}, we can perform an additional masking of the telluric lines with absorption deeper than 70\% of the continuum level\footnote{This mask extends from the line center until the relative absorption is lower than 5\% of the continuum level.}. Diffraction orders~57 to~54 (i.e. $\sim$1\,300 to $\sim$1\,500\,nm) and 42 to 40 (i.e., $\sim$1\,800 to $\sim$2\,000\,nm), located within nIR water absorption bands, are discarded in what follows\footnote{Due to their significant fraction of high-absorption / saturated telluric lines, keeping these orders in the analysis leads systematically to degraded results.}. We use a data analysis that resembles that of \cite{Boucher2021} albeit with a few differences described below. Our analysis consists of the following steps, independently applied to each of the 42 remaining diffraction orders, and illustrated on a given order in Fig.~\ref{fig:analysis}.

\begin{enumerate}
    \item We create an out-of-transit stellar reference spectrum, $\boldsymbol{I_{\rm{ref}}}$, by averaging the out-of-transit spectra interpolated in the stellar rest frame (the star moves by $\sim$200 m.s$^{-1}$ during the course of the transit). This reference spectrum, shifted back to the Earth rest frame, is linearly adjusted in flux to each observed spectrum $\boldsymbol{I_{\rm{obs}}}$, and $\boldsymbol{I_{\rm{obs}}}$
    is divided by this best-fitting solution. This operation is then performed once again to the resulting spectra (hereafter reference-corrected spectra), but, this time, the out-of-transit reference spectrum is computed in the Earth rest frame in order to remove residual telluric contamination. Note that, contrary to \citet{Boucher2021}, the rest of our data analysis is conducted in the Earth rest frame rather than in the stellar rest frame, so that the interpolation of the noise only affects the master (median) spectra and not individual spectra. Additionally, note that our data reduction enables the user to correct for planet-induced distortions of the stellar line profiles \citep[e.g., center-to-limb variations or the Rossiter-Maclaughlin effect; see][]{chiavassa2019}. At each epoch, the code linearly fits a user-provided distorted stellar spectrum to the data prior to step~(i) and normalise the data. As planet-induced distortions of the stellar line profiles are not taken into account in our simulations, we will not give further details on its implementation in this paper and redirect the reader to a paper in preparation (Klein et al., in prep). Note that, as shown in the panel~[b] of Figure~\ref{fig:analysis}, low-frequency variations in time and wavelength domains are still identifiable after correcting for the reference spectrum. These residual variations are most likely due to modal noise from the fibers \citep{oliva2019} and requires additional normalisation.

    \item  We normalise each residual spectrum by a noise-free continuum estimate, computed using a rolling mean window, and remove outliers in the normalised spectra using a 5-$\sigma$ clipping procedure. These two steps are repeated until no outlier is flagged in the data. By measuring how the variance of the normalised spectra varies with the size of the rolling window, we find that a minimum width of $\sim$23\,\kms (10 SPIRou pixels) is required to reliably average the spectrum noise. The exact size of the window has no more than a marginal impact on the recovered planet signature, provided that it is small-enough to correct for the low-frequency structures induced by modal noise in the data. In what follows, we fix the window size to 50 pixels (115 \kms) and discard the same amount of points at the ends of each diffraction order.

    \item At this stage, outliers have been removed in the wavelength space, for each spectrum individually. However, some pixels (i.e. wavelength bins) might still exhibit large temporal variations (e.g. due to telluric contamination). In order to flag and remove these high-variance pixels, we compute the variance in the wavelength space (i.e. for each pixel), and perform an iterative parabolic fit with a 5-$\sigma$ outlier removal to the variance distribution\footnote{Note that, as a result of the blazed grating, the noise at the edge of each diffraction order is larger than in the center, which justifies the choice of a parabolic fit}. The outliers flagged in the process are masked out in the subsequent steps of the data processing.

    \item Our data processing pipeline features an optional additional filtering step to correct for the variation of residual telluric absorption with airmass. Accordingly to \citet{brogi2018}, we fit a second order polynomial of the log of the intensity with airmass and remove it out (hence divide it out in intensity). However, the airmass is no more than an incomplete proxy of the water telluric absorption, expected to unpredictably change 
    on short time scales during the observations. As a consequence, several studies have chosen to bypass this step \citep[e.g.,][]{dekok13,Boucher2021,Giacobbe2021} in favour of a more statistical approach (often based on PCA). In this study, we have kept the quadratic airmass detrending but it can be easily by-passed in our pipeline (and the order of the polynomial can also straightforwardly be changed).

\end{enumerate}

\begin{figure*}
    \centering
    \includegraphics[width=\linewidth]{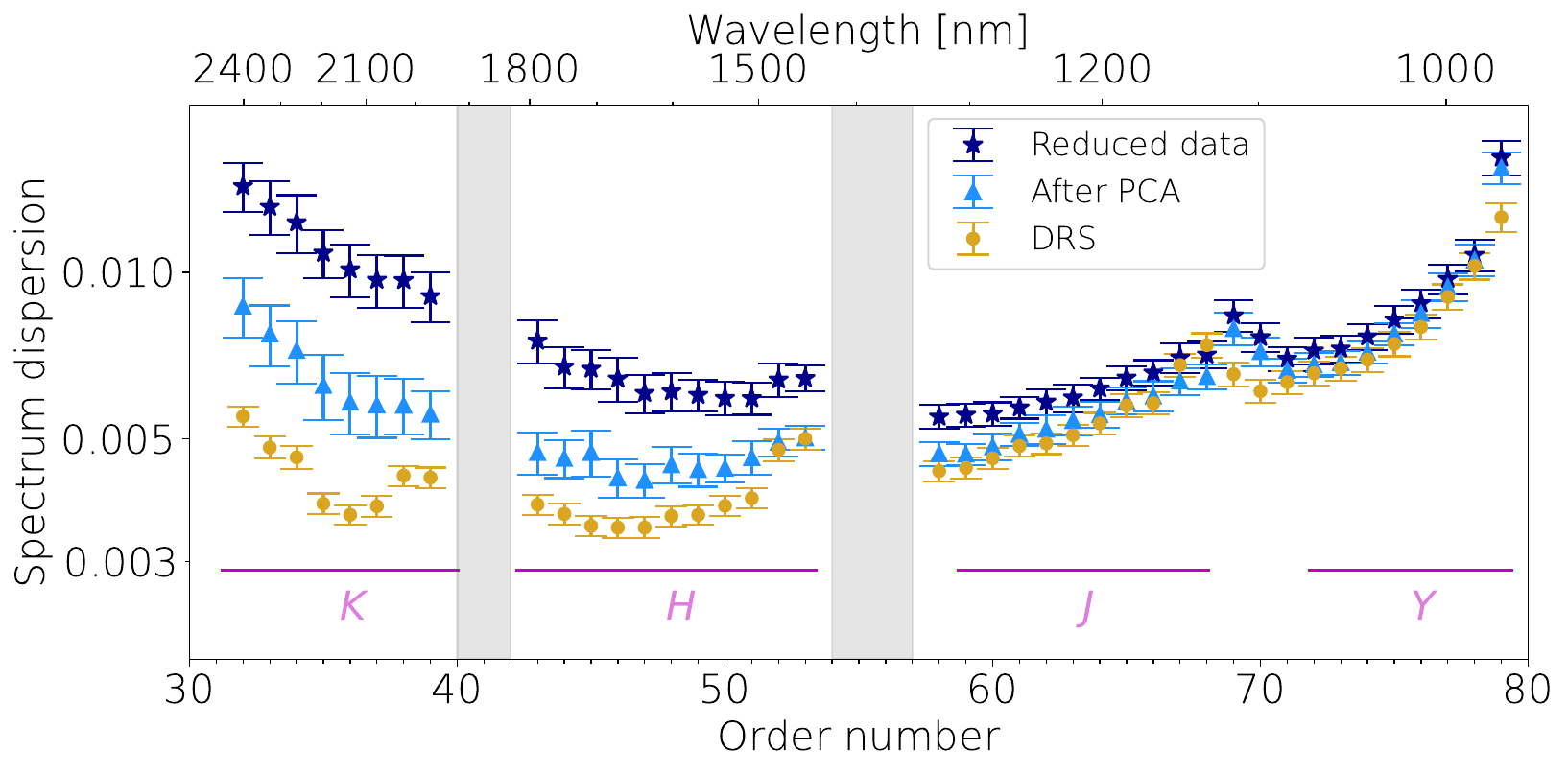}
    \caption{Variance at the center of each diffraction order after processing the data as described in Section~\ref{ssec:analysis} before and after the PCA (or auto-encoder) reduction (resp. dark-blue stars and light-blue triangles). As a reference, the photon noise estimate provided by APERO is shown in gold filled circles. Each point and error bars give the mean and standard deviation across all spectra. Orders removed due to strong telluric contamination are indicated by the vertical gray bands. The position of the $YJHK$ photometric bands is indicated by the horizontal magenda solid lines.}
    \label{fig:dispersion_reduc}
\end{figure*}

The distribution of the variance in wavelength of the processed spectra is compared to the APERO-provided photon noise in Fig.~\ref{fig:dispersion_reduc}. The dispersion of the processed spectra remains similar to the APERO estimates for the blue half of the spectrum, but is significantly higher in the $H$ and $K$ bands. This is most likely due to the fact that modal and thermal noises, stronger at redder wavelengths, are not including in the APERO estimation of the photon noise. To correct for residual correlated noise due to both imperfect corrections of the stellar and Earth absorption spectra and instrumental noise, we apply an additional data-driven procedure describe in the following section.

\subsection{PCA and Auto-encoder}\label{ssec:pca}

The last step of the data processing is conceptually different to the others in the sense that we aim at getting rid of the correlated variance in time in our data on which we have no physical priors. Defining a deterministic framework to do so seems impracticable as we expect this correlated noise to highly depend on the target and on the observing conditions. This step is therefore necessarily a data-driven approach. In practice, we have developed two different methods that we independently apply to the data. Having two methods to statistically reduce the correlated noise in the data provides additional robustness to any claim of planet atmosphere detection and prevents false positives. We stress that both methods are applied to the log of the reduced data where we can consider at first order that the total spectra is a linear combination of the planet's and noise. 

Our first method is based on principal component analysis (PCA). PCA is a linear method, that recovers the dominant source of correlated variance from an eigenvector decomposition of the covariance matrix: the principal components are these eigenvectors sorted by decreasing eigenvalues. This technique has been extensively used and discussed in several HRS-based planet atmosphere studies \citep[e.g,][]{dekok13,damiano2019,Brogi2019,Boucher2021,Pelletier2021}. Our second method uses a deep-learning approach based on an auto-encoder, and is a new method in the HRS exoplanet community. An auto-encoder is an artificial neural network which aims at reproducing the dominant features of a data set by encoding them into a much lower number of points through subsequent reduction matrices. Initially proposed by \citet{Hinton2006}, it is now widely used in many fields of applied mathematics and in some astrophysical works as well \citep[e.g.,][]{Yang2015,Cotar2021}. In essence, both methods rely on reducing dimensionality by transforming our spectra into smaller sets, but, unlike PCA, our auto-encoder is not linear and we lose information about how the data is coded\footnote{See this comparison between the PCA and auto-encoder: \url{https://towardsdatascience.com/autoencoders-vs-pca-when-to-use-which-73de063f5d7}.}. We now give details on the practical implementation of boths methods in the next two paragraphs.


\subsubsection{PCA implementation}

\begin{figure}
    \centering
    \includegraphics[width=\linewidth]{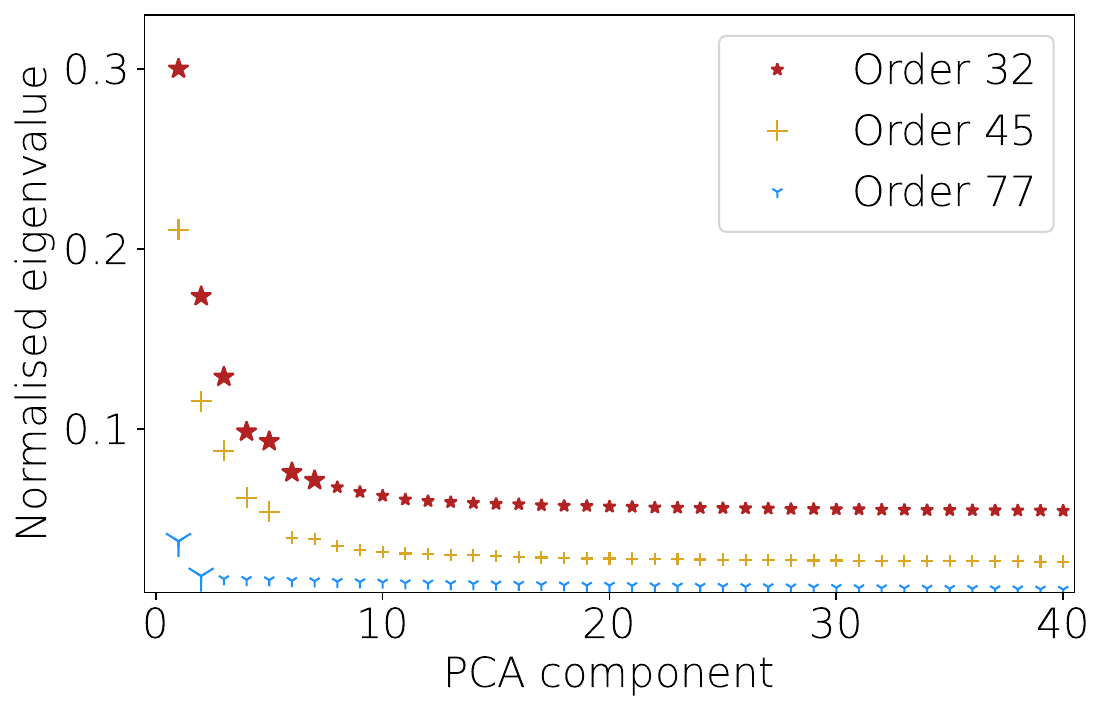}
    \caption{Eigenvalues associated to each PCA component of SPIRou \'echelle orders 77 ($\sim$999~nm, light-blue symbolds), 45 ($\sim$1\,707~nm, gold crosses) and 32 ($\sim$2\,400~nm, dark-red stars). The eigenvalues for each of the three orders have been vertically shifted for better clarity. For each order, larger symbols indicate components which have been flagged, in Sec.~\ref{ssec:pca}, as dominated by correlated noise and removed in the analysis, i.e. 2, 5 and 7 components for orders 77, 45 and 32, respectively.}
    \label{fig:pca_sel}
\end{figure}

In our data analysis pipeline, the PCA-based dimensional reduction is applied independently to each order in the time domain\footnote{The structure of the correlated noise is expected to vary from one order to the next and, therefore, the number of components associated with correlated noise and subsequently discarded has no reason to be the same for all orders.}. The number of components associated with correlated noise and subsequently discarded for the analysis is tuned using the following procedure, illustrated on a given order in Fig.~\ref{fig:pca_sel}. For each order, we generate 5 to 10 sequences of spectra matching our observed wavelengths and times, but containing only uncorrelated Gaussian noise of level similar to our empirical photon noise estimate. To account for the larger noise level at the edges of the order, the sequences of noise are amplified by the normalised inverse of the square root of the blaze function. In principle, these sequences are free from correlated noise and can be used as references to tune our PCA. We apply PCA to each sequence of noise and store the largest eigenvalue, $S_{\rm{max}}$. When we apply PCA to our observed sequences of processed spectra, any component associated with an eigenvalue significantly larger than $S_{\rm{max}}$ (e.g. $2 \times S_{\rm{max}}$, see the red dotted line in Fig.~\ref{fig:pca_sel}) likely encloses a significant amount of correlated noise, and is discarded. For \GLA, this procedure typically removes 4 PC in the blue part of the spectrum and 8 in the reddest part, which we attribute to the complex structure of the stellar atmosphere and/or modal noise. For the hotter star HD 189733 (see section \ref{sec:HD189}), we typically remove 2 components in the blue part and 5 to 7 in the red part.
Note that injecting the synthetic planet signature to the noise maps has a (i)~marginal impact on the eigenvalues and (ii)~affects all the components by more-or-less the same factor, ensuring the planet atmosphere spectrum is not removed in the process. This effect is further discussed in Section~\ref{ssec:degrading}. Finally, note that the weighted PCA framework of \citet{delchambre2014}, is also implemented in our publicly-available data processing code (via the \texttt{wpca} python module).

\subsubsection{Auto-encoder implementation}

 As in \citet{Cotar2021}, our implementation of the auto-encoder relies on 4 layers, which allows us to reduce the data dimensionality from a few thousands pixels (the size of a corrected SPIRou order, which varies from 2000 to 4000 pixels after removal of bad pixels from telluric correction) to 8. Each SPIRou order has a different auto-encoding process, but, in the same order, the spectra have a common encoding and decoding matrix: as for PCA, the auto-encoder takes into account time correlated features. We train the network in the following way: we randomly select 70\% of the reduced spectra (panel 3 of Fig. \ref{fig:analysis}), encode them through the four layers which reduce dimensionality (to 1024, then 256, then 64 then 8~pixels) and then reconstruct the original spectra. All these numbers are just optimisation of the auto-encoding process, and can of course be changed (the number of layers as well). This creates an auto-encoder which is then applied on the 30\% remaining spectra in order to validate that their reconstruction is reliable as well. After 5000 iterations, we consider that the network has sufficiently learned and use the resulting encoding matrix as our final neural network. We then apply the final auto-encoder on the reduced spectra to create a reconstruction of the dominant feature and remove it from the observations. This is equivalent to the way we apply PCA, although we completely lose the information about the number of components order by order and how they are encoded. 
 
 The auto-encoder takes much longer to run than PCA (typically $1$min per order on a GPU against $0.1$s on a CPU for PCA) because of the learning curve. However, once the algorithm has learned and its transformation matrices are defined for a given sequence of spectra, it takes only a few milliseconds to run it on a CPU.

\subsection{Uncovering the planetary signature}
\subsubsection{Template matching}\label{ssec:ccf}

Once the reduced data have been cleaned through the PCA or auto-encoder, the planetary signal is still largely buried under the noise as can be seen on the last panel of Figure \ref{fig:analysis}. The use of a correlation function between a theoretical model and the reduced data has therefore been proposed since the first successful exoplanet atmosphere characterisation by HRS of \citet{Snellen2010}. As is done in the literature, we first create an atmospheric model at extremely high spectral resolution (between 300 000 and 1 million) with \texttt{petitRADTRANS}. We then use this model to build a sequence of synthetic spectra matching the observing epochs and wavelengths. This requires to Doppler-shift the model by the planet RV, $V_{\rm{p}}$, computed at each observed planet orbital phase $\phi$ using 

\begin{equation}
V_{\mathrm{p}} \left( \phi \right) =  K_\mathrm{p} \sin \left( 2 \pi \phi \right) + V_\mathrm{sys} \text{,} 
\end{equation}

\noindent
for different values of the planet velocimetric semi-amplitude (\kp) and systemic Doppler shift (\vsys), and convoluting the shifted models with a Gaussian at SPIRou's resolving power. The synthetic sequence is then processed with some of the key steps described in the previous sections, as we expect the data analysis to affect the planetary spectra. This is described in Section \ref{ssec:degrading}.

Finally, we build sequences of processed synthetic spectra for a range of \kp\ and \vsys\ values, and compute the scalar product between each of these sequences and the observed spectra to create a correlation function (as in \citet{Boucher2021}):
\begin{equation}
    CCF = \sum_i \left. \dfrac{d_i m_i}{\sigma_i^2} \text{,} \right.
\end{equation}
where $d_i$, $m_i$ and $\sigma_i$ are respectively the observed flux, the model value and the flux uncertainty at pixel $i$ (corresponding to time $t$ and wavelength $\lambda$).
Our correlation maps typically extend from K$_p$\,=\,0\,\kms\ to twice the theoretical value of \kp, computed from the masses of the star and planet and the semi-amplitude of the planet-induced stellar RV wobble. For \vsys, we typically explore a 200 \kms\ wide window centered on 0. A detection can be claimed if the maximum of correlation between the reduced data and the model is obtained close to the injected semi-amplitude and Doppler shift. Following \citet{Boucher2021}, we define $\sigma_i$ as the standard deviation of the value of the pixel $i$ weighted by the S/N of each spectrum:

\begin{equation}
    \sigma_i^2 = \sigma^2 (t,\lambda) = \dfrac{\sum_t \left( d(t,\lambda)-\overline{d(\lambda)}\right)^2}{N_\mathrm{spectra}}\dfrac{\overline{\mathrm{SNR}}}{\mathrm{SNR}(t)}
    \label{eq:sigma}
\end{equation}
where the bar denotes a time average, $N_\mathrm{spectra}$ is the number of spectra and $d_i = d(t,\lambda)$.
Finally, in order to convert the correlation values to significance of detection, we divide the former by the standard deviation of the correlation map in regions dominated by white noise (i.e. away from the planetary signal), as frequently done in the literature. Note that the cross-correlation analysis is only used for first-order searches of planet signatures, whereas a more statistically robust (but more time consuming) exploration of the parameter space is performed in the Bayesian framework described in Section~\ref{ssec:MCMC-nest}.

In terms of speed, we tried to optimise the calculation of this correlation in the public code, and for a low resolution map (50 $\times$ 50 points in $K_p$ and $V_\mathrm{sys}$), it typically takes a couple of minutes per transit over the whole SPIRou domain on one processor. We have not parallelized it as this is sufficiently efficient for the use we make of it, but it would be very straightforward to do so by splitting the calculations for different regions of the (\kp,\vsys) map.

\subsubsection{MCMC and nested sampling}
\label{ssec:MCMC-nest}

Finally, we have the possibility to robustly explore the parameter space in a Bayesian framework. We implemented two methods: a Markov Chain Monte Carlo algorithm (MCMC) based on the python module emcee \citep{Foreman2013} and a nested sampling algorithm based on the python module \texttt{pymultinest} \citep{Buchner2014,Feroz2008,Feroz2009,Feroz2019}. The second one is typically 50 times faster than the first one, and, in our tests, we never noticed any difference in the results of these two methods. We therefore present only results using the nested sampling algorithm in the rest of the paper, but having the possibility to use both samplers allows us to have independent avenues to validate the results. Both methods rely on the a likelihood $\mathcal{L}$, defined in \citet{Brogi2019} and \citet{Gibson2020} by

\begin{equation}
    \mathcal{L} = \prod_{i} \dfrac{1}{\sqrt{2 \pi \zeta_i}} \mathrm{exp} \left\{ -\dfrac{[am_i- d_i]^2}{b \zeta_i^2} \right\} \text{,}
    \label{eq:like}
\end{equation}
where $\zeta_i$ accounts for the uncertainty of the i$^{th}$ pixel and $a$ and $b$ are scaling factors to account for incorrect modelling and incorrect estimation of the order variance, respectively. In the rest of this paper, a is set to 1. The main difference between the two approaches is that \citet{Brogi2019} uses a unique $\zeta$ value that does not depend on the pixel. They derive the likelihood relative to $\zeta$ and select the value that cancels the derivative, hence ensuring a maximum of likelihood , whereas \citet{Gibson2020} allows $\zeta$ to be defined pixel by pixel. 

When applying the \citet{Brogi2019} likelihood, $b$ is set to 1 and we calculate the optimal $zeta$ for each spectrum (which is what the authors advise (M. Brogi, private comm.)). Essentially, this is similar to say that our log-likelihood is the sum of the log-likelihoods of each spectrum, with a different $\zeta$ optimised for each spectra. On the other hand, with the Gibson likelihood, the values of $\zeta_i$ are defined from prior information: in this paper we chose $\zeta_i = \sigma_i$, as defined in Eq.\eqref{eq:sigma} The $b$ value is then obtained in a similar manner than the $\zeta$ in \citet{Brogi2019}: we chose $b$ that cancels the derivative of the likelihood, as explained in \citet{Gibson2020}. We have the freedom to optimise this $b$ value for (i)~each spectrum, (ii)~each order, (iii)~each transit, or (iv)~globally. We found that, for the simple tests presented in this paper, the four options provided very similar results.

The typical time to converge a nested sampling algorithm for a model with 4 parameters ($K_p$, $V_\mathrm{sys}$, temperature and water mass mixing ratio) and 384 live points, which are the nested sampling equivalent of a walker in a usual MCMC framework, is 2-3 hours on 36 processors. This gets drastically longer with more molecules as we face a memory issue, which is inherent to petitRADTRANS for now (P. Molliere, private discussion). This problem could be overcome by precomputing a grid of models and interpolating in this grid rather than calculating a model at each iteration (which is our choice here), but that becomes prohibitively complicated with too large a number of molecules (typically $\gtrsim 3$).

\subsubsection{Degrading the model}
\label{ssec:degrading}

Although the PCA or the auto-encoder mainly remove planet-unrelated noise, they do affect the planetary signature in the data. There is a lot of work in the literature to reproduce at best the degradation by PCA onto the synthetic model so as to optimise the template matching function and/or likelihood calculations \citep{Brogi2019,Gibson2020,Boucher2021,Pelletier2021}. Skipping this step leads to significant errors in the retrieved atmosphere parameters (see the discussion in \ref{ssec:disc-degrade}).

\begin{figure}
    \centering
    \includegraphics[width=\linewidth]{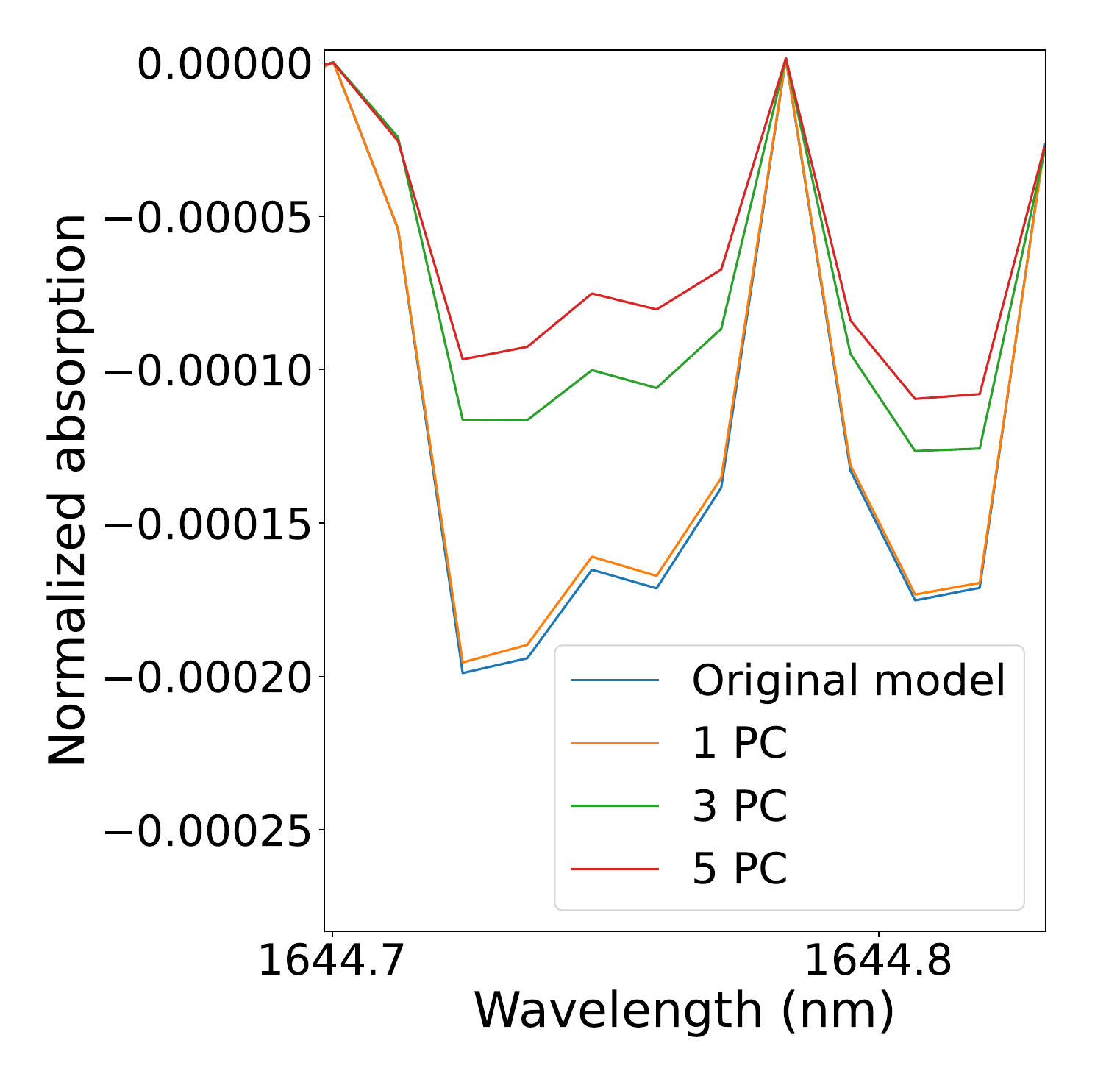}
    \caption{Zoom on a planet atmosphere model degraded with \citet{Gibson2022}'s framework (see Section~\ref{ssec:degrading}) when removing 0, 1, 3 or 5 principal components (PC). }
    \label{fig:degraded}
\end{figure}

In order to be performed in the nested sampling algorithm, this degradation must be as fast as possible numerically. We have therefore implemented the fastest of these methods for PCA (i.e. that of \citet{Gibson2022}), which we detail below. We have not yet found an equivalently fast method for the auto-encoder, because of non linearities in the process, and, therefore, we limit the statistical exploration of the parameter space to PCA-reduced data. However, using PCA we have realised that {\it{not}} degrading the model was not an issue for molecular detection when performing template matching: it only reduces the significance marginally. Our use of auto-encoder can therefore be applied to molecular detection through template matching, but is not yet suited for parameter retrieval.

Our implementation of the \citet{Gibson2022} method for PCA works as follows. During the data reduction, we store the removed eigenvectors (i.e. associated with correlated noise) for each order into a matrix $\mathrm{U}$ (calculated in log-space). We then multiply $\mathrm{U}$ by its pseudo-inverse $\mathrm{U}^\dagger$ to create an orthogonal projector of the vector space defined by these eigenvectors. We then project the logarithm of our synthetic sequence model $\mathrm{M}$ on this vector space, and remove it from $\mathrm{M}$. Our final degraded sequence $\mathrm{M}'$ is therefore given by


\begin{equation}
    M' = \exp \left(\log M - U U^\dagger \log M \right) \text{,}
\end{equation}

\noindent
We stress again that $U$ changes from one order to the other. As in \citet{Gibson2022}, we do not need to take the weights into account as they can be naturally implemented in the weighted PCA algorithm. Fig.~\ref{fig:degraded} shows the effect of such a degradation on an isothermal model of our HD189733\,b-analog, containing only water, when 1, 3 and 5 PCA components are removed of the data (for order 52 here).

\subsection{Including rotation and winds}

\citet{Brogi2016} defined a framework to include the effect of rotation and winds on a 1D transmission spectra in a phase dependent manner. This method is however quite time consuming and hence not suitable for a large parameter space exploration. In Appendix~\ref{app:winds}, we show that the inclusion of rotation can be expressed as a double convolution at mid transit. This provides a very fast first-order calculation of the effects of rotation (and eventually winds, see the appendix), albeit not as accurate as the framework of \citet{Brogi2016}, since it does not take the phase dependence or limb darkening effects into account. In what follows, we rely on the equations presented in Appendix~\ref{app:winds} to recover planetary rotation at first order in our nested sampling algorithm. In particular, note that by separating Eq.~\ref{eq:first_eff_rad} into its blueshifted and redshifted components, one could straightforwardly create a transmission spectrum where both hemisphere have different physical parameters. This hemispheric dichotomy is notably applied in a forthcoming work of the ATMOSPHERIX consortium (Hood et al., in prep.).

\section{Application on simulated data}\label{sec:Results}

\subsection{Simple isothermal model}\label{ssec:simple_model}

Following the process described in Section~\ref{sec:planet_injection}, we first inject a simple, isothermal atmospheric model, containing only water with a volume mixing ratio of $10^{-2}$ and a temperature of $900$K, in the APERO-provided telluric-corrected spectra of \GLA. As a first step, we run the cross-correlation analysis to the data processed using the different steps described in Sec.~\ref{ssec:analysis}, but prior PCA (or auto-encoder) cleaning. Unsurprisingly, strong signatures at low semi-amplitudes and velocity shifts dominate the correlation map, confirming that residuals water lines from \GLA\ and the Earth atmosphere are still prominent in the reduced spectra. Note that detrending the data with airmass is not sufficient to uncover the injected signal, which confirms that a PCA / auto-encoder treatment is needed.

\begin{figure}
    \centering
    \includegraphics[width=\linewidth]{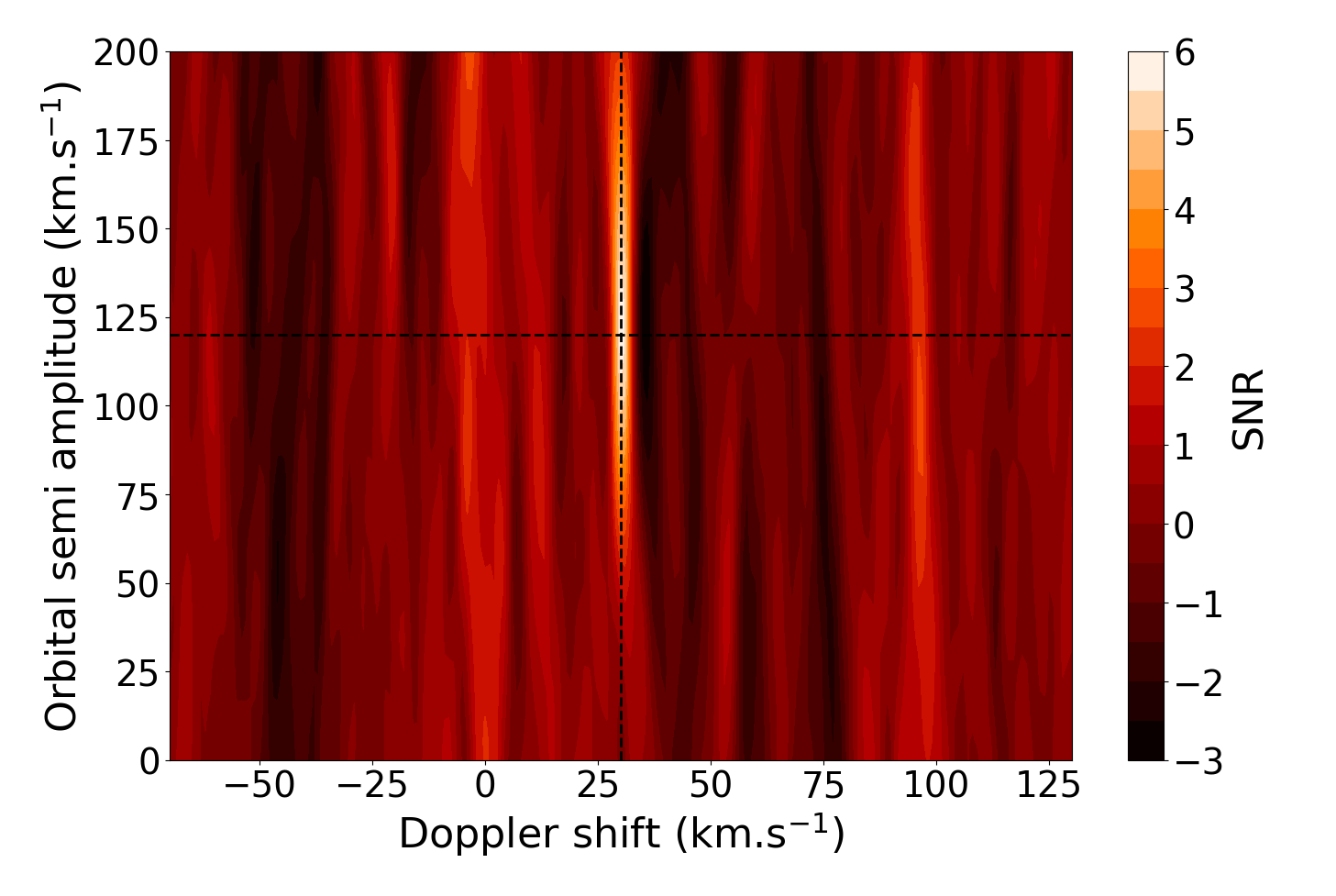}
    \caption{Cross-correlation significance as a function of Doppler velocity and orbital semi-amplitude for the sequence of \GLA\ spectra, in which an isothermal planet atmosphere model has been injected (see Section~\ref{ssec:simple_model}), and reduced using the procedure described in Section~\ref{ssec:analysis} and PCA-cleaning (Section~\ref{ssec:pca})}.
    \label{fig:correl_PCA}
\end{figure}

\begin{figure}
    \centering
    \includegraphics[width=\linewidth]{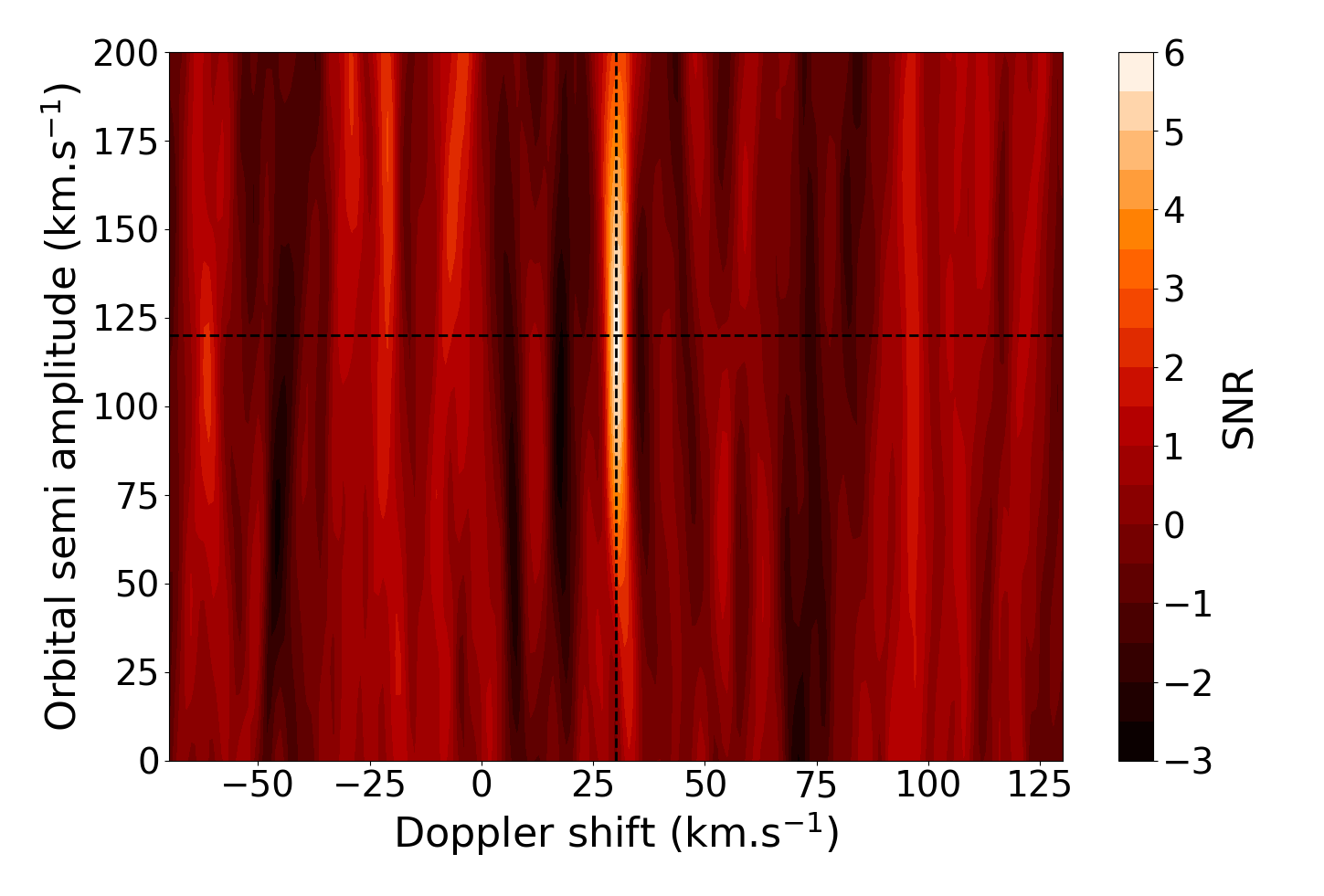}
    \caption{Same as Fig. \ref{fig:correl_PCA}, this time for the encoder-reduced data.}
    \label{fig:correl_enco}
\end{figure}

In contrast, fair detections of the injected signals are obtained when the data are cleaned with PCA or auto-encoder. In terms of cross-correlation, the signal was found at a signal-to-noise ratio of about 6, as shown in Fig.~\ref{fig:correl_PCA} and~\ref{fig:correl_enco}. Both the auto-encoder- and the PCA-based treatments yield similar level of signal detection, thereby confirming that the auto-encoder is a reliable, robust approach to PCA, which would gain at being developed further. The relatively large significance of detection \citep[of the order of what can be found on hot Jupiters; see][]{Line2021} can be explained by the fact that the planet absorption template used in the cross-correlation is the same as the injected model.


\begin{figure*}
    \centering
    \includegraphics[width=\linewidth]{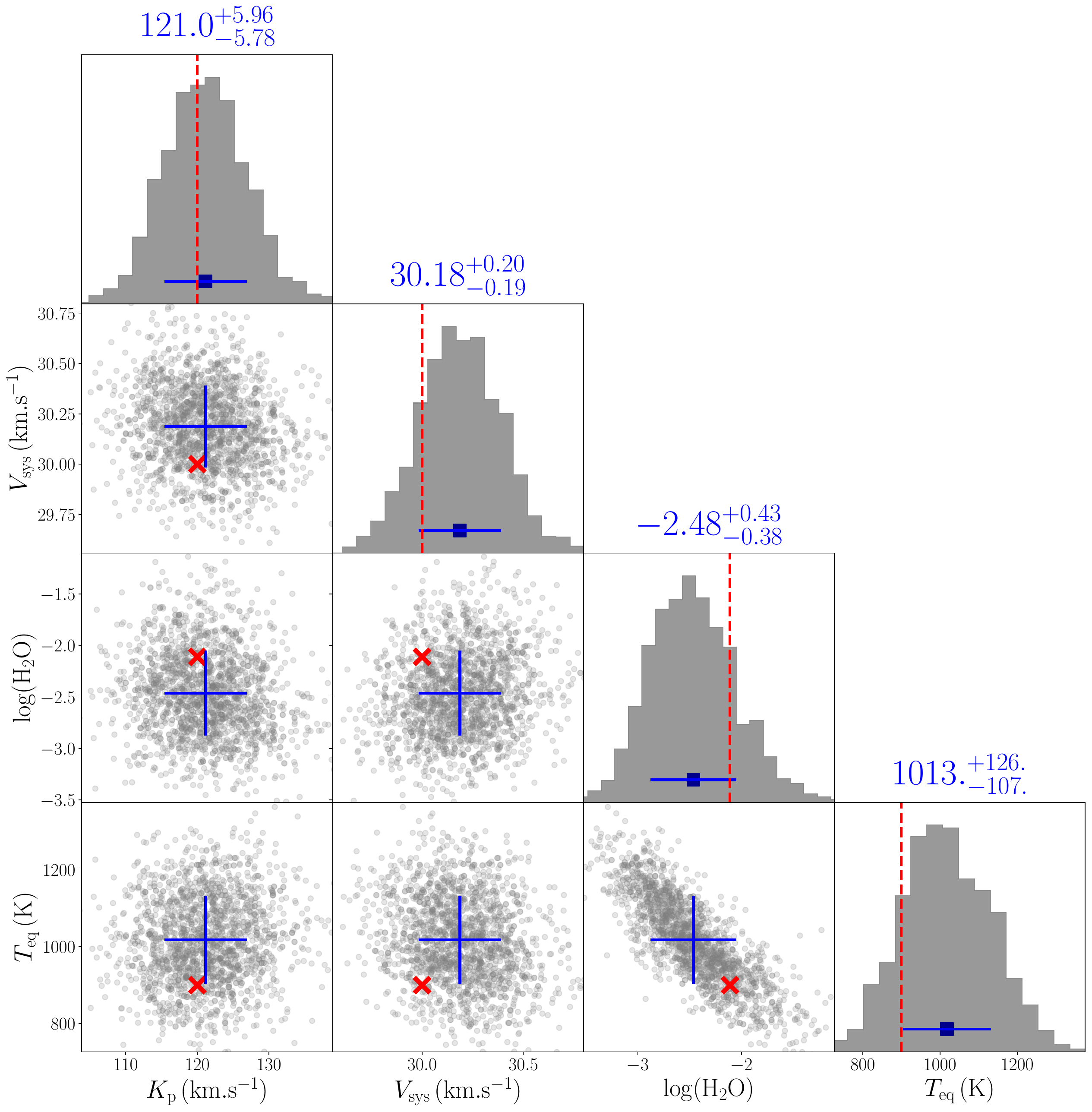}
    \caption{Posterior densities resulting from a \texttt{pymultinest} retrieval of the datacontaining the injected planet atmosphere signal described in Section~\ref{ssec:simple_model}, with the simple isothermal model described in Section~\ref{ssec:simple_model}. The blue squares with error bars show the best-fit values, with the $\pm$\,$1\sigma$ uncertainties, and the red line and crosses indicate the injected values }
    \label{fig:nested_simple}
\end{figure*}

In terms of parameter retrieval from the nested sampling algorithm, Fig. \ref{fig:nested_simple} shows the corner plot of our results. Our MCMC process has converged towards Gaussian-looking posterior densities, roughly matching the injected parameters. Temperature and water content are unsurprisingly degenerated, but we do recover the injected values in the 1-$\sigma$ ellipse of posteriors. Note that changing the white noise realisation (by fitting the other synthetic transit) or the likelihood definition (see Section~\ref{ssec:MCMC-nest}) only marginally affects the retrieved parameters, which remain consistent within $\sim$1$\sigma$. This confirms that our analysis and parameter estimation process do not introduce strong biases in the retrieval. Additionally, we tested the effects of changes in the input water content and did not find systematic trends in the parameters recovered using different likelihood definitions: a same likelihood can overestimate the water content for one synthetic model and underestimate it for another model. On real data, we therefore recommend to gather results from different likelihoods to define conservative error bars in the retrieved parameters.

\subsection{Including rotation}

\label{ssec:rotation}

\begin{figure*}
    \centering
    \includegraphics[width=.6\linewidth]{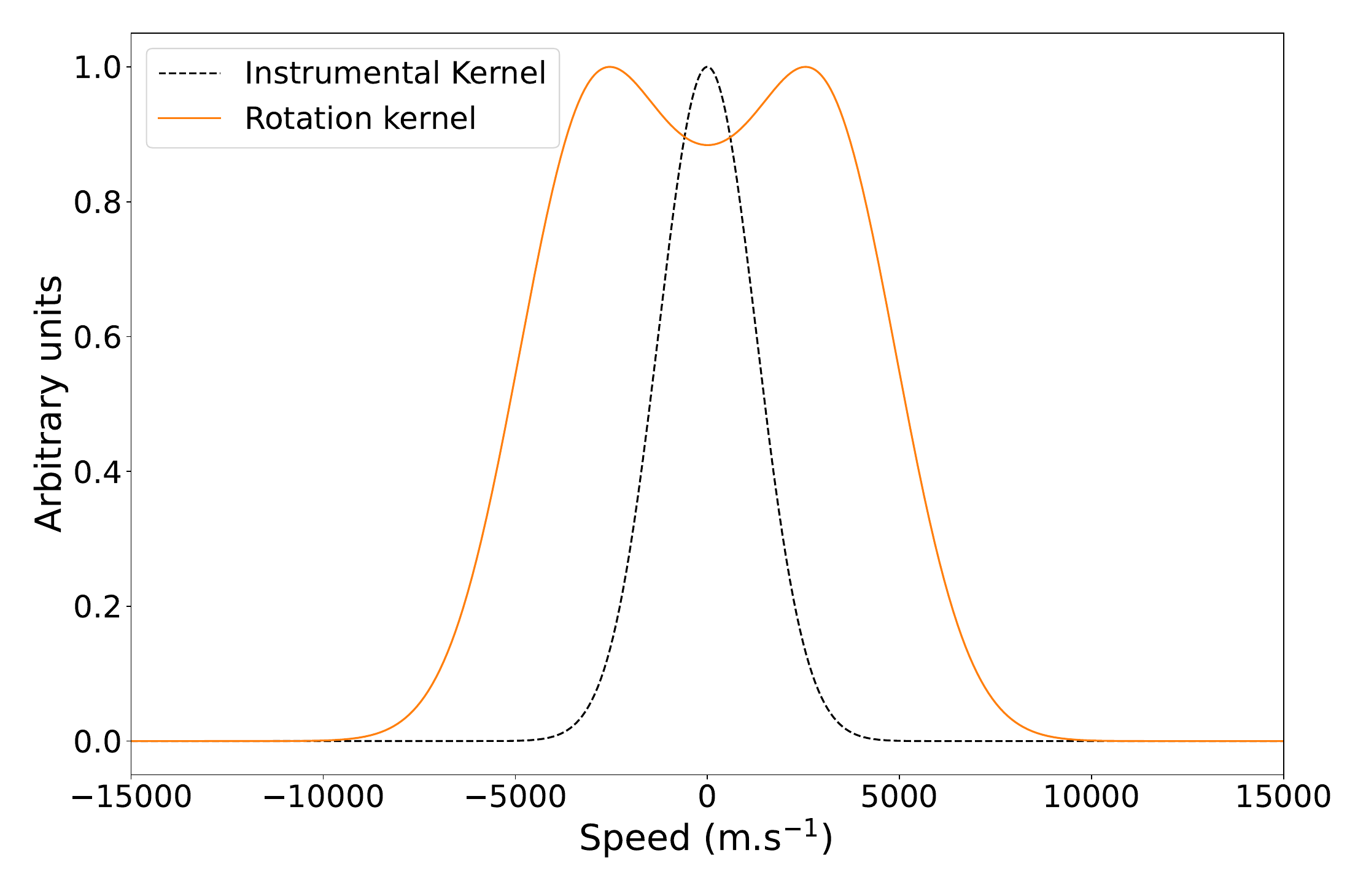}
    \includegraphics[width=.6\linewidth]{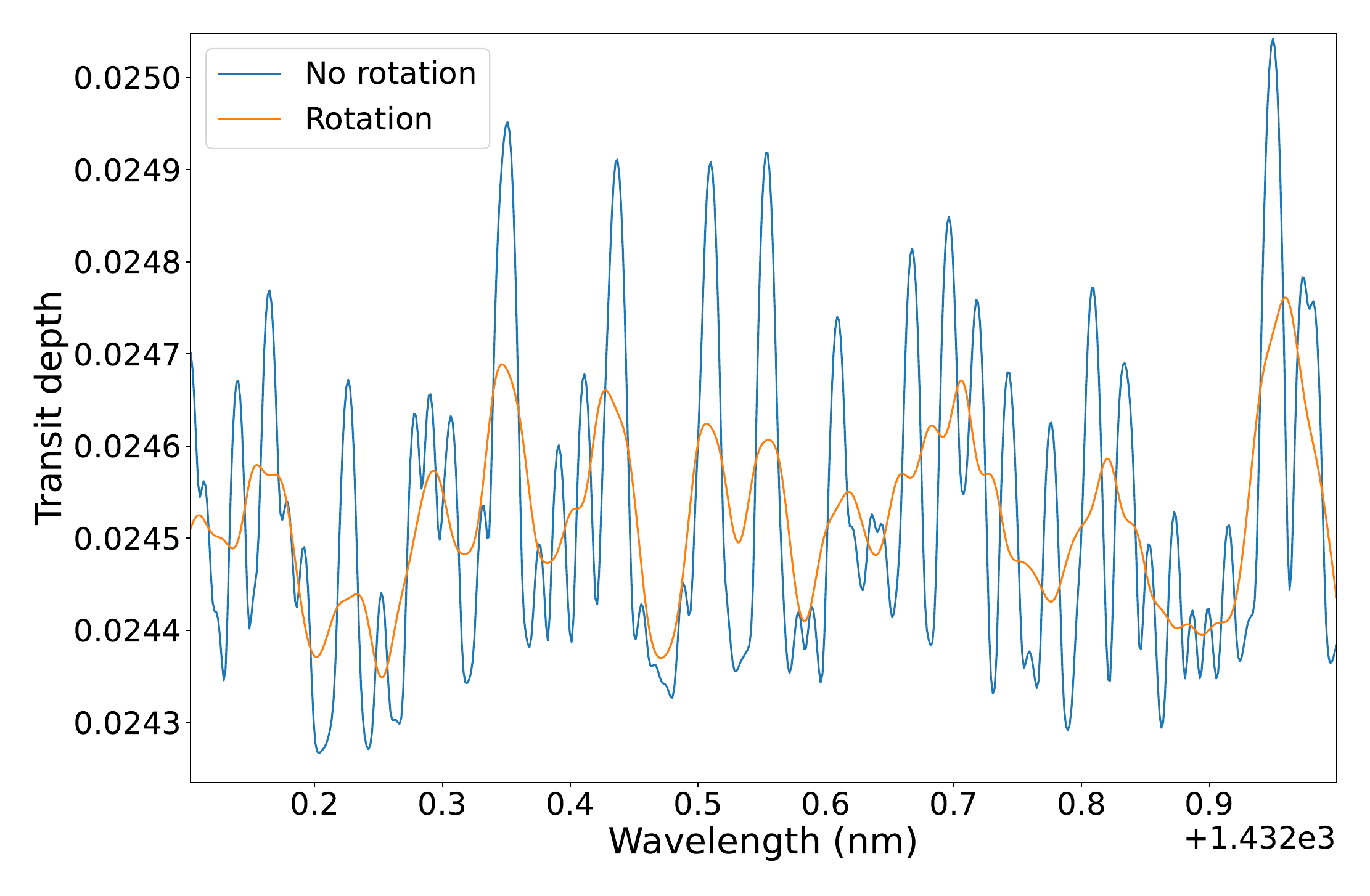}
    \caption{Top: rotation kernel (arbitrary units) in plain orange line and instrumental profile in dashed line as a function of velocity similar to \citet{Brogi2016}. Bottom: zoom on the absorption lines of a model containing only water, broadened at SPIRou resolution, and the same model when applying a rotation with an equatorial velocity of 3000\,\ms.}
    \label{fig:model_rotated}
\end{figure*}

We have first tried to recover rotation from a model that did not include any, and we found that our nested sampling algorithm could not differentiate between a non rotated model or models with equatorial rotation lower than than 1 \kms. Then we have created a model with a planetary rotation of $3$\,\kms\ at the equator, following appendix \ref{app:winds} (hence assuming the rotation axis is perpendicular to the orbit). When we try to recover this model with a non rotating model, the best-fit value of the water mixing ratio is decreased by a factor of $\sim$\,30 and we also get lower values for the temperature. This is due to the fact that rotation decreases the strength of the absorption lines by spreading them on a larger width (see Fig.~\ref{fig:model_rotated}), whereas temperature and water content typically increase the line contrasts. As we are mostly sensitive to line amplitude and not shape, this creates a degeneracy between rotation and temperature/composition.

\begin{figure*}
    \centering
    \includegraphics[width=\linewidth]{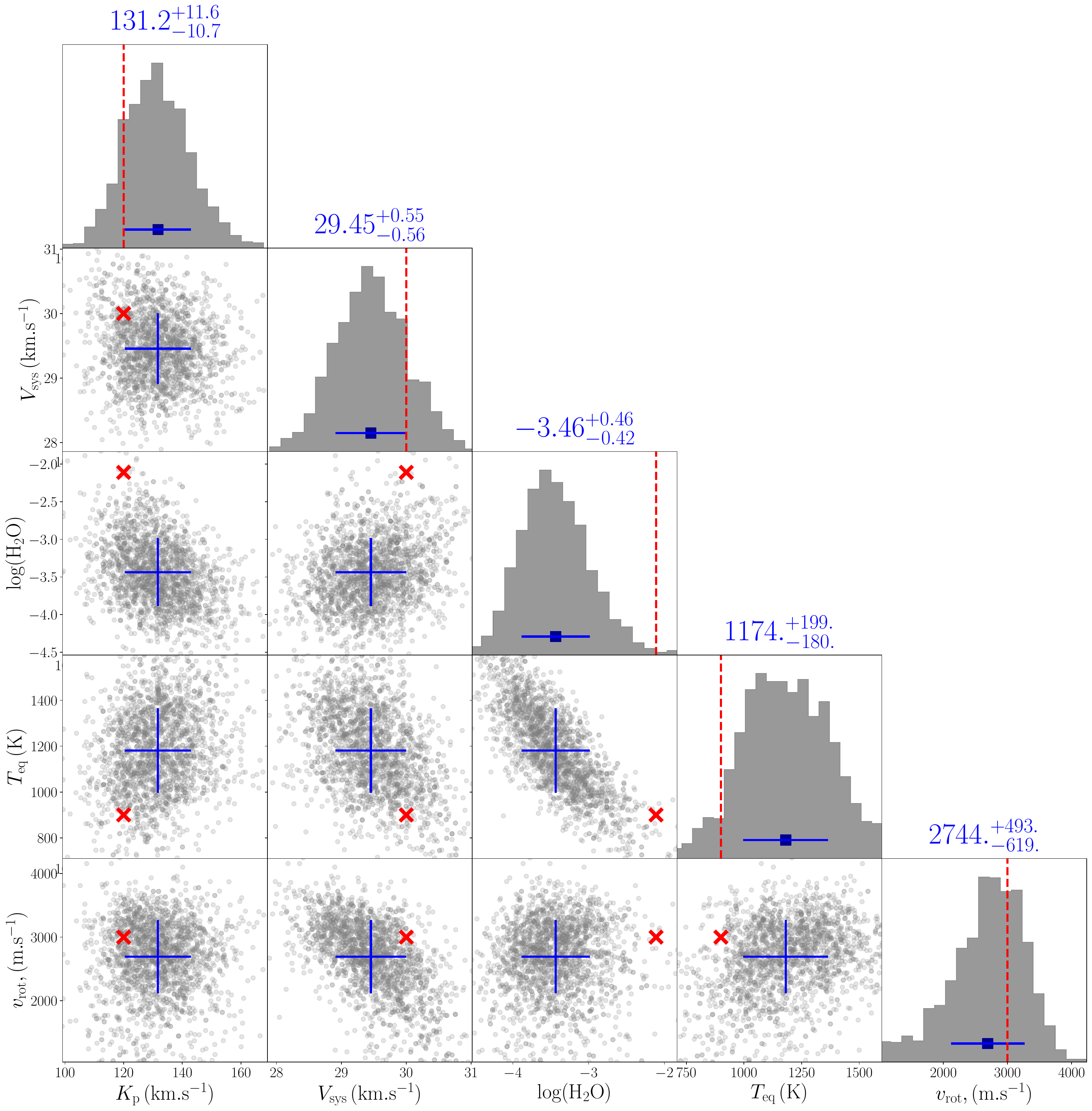}
    \caption{Corner plot showing the results of a \texttt{pymultinest} retrieval with a model containing only water with a mass mixing ratio of $10^{-2}$ and a rotation with equatorial velocity of 3\,\kms. The different figure elements are the same as in Fig.~\ref{fig:nested_simple}.}
    \label{fig:marg_rotated}
\end{figure*}

When we try to retrieve a model with rotation, we recover the correct parameters in the posteriors as shown in Fig.\ref{fig:marg_rotated} but the mean recovered values for water and temperature are 3 and 2$\sigma$ away from the maximum of posterior probability for water and temperature. No matter the rotation speed, we always obtained too low water content and too high temperature, showing that this is intrinsic to the analysis which exhibits a bias for large rotation rate. As demonstrated in Appendix \ref{app:bias}, we have performed a wide range of tests and demonstrated that it most likely comes from a lack of model degradation. Indeed, although our model is degraded consistently with the PCA applied to the data, the first phases of the data analysis (average stellar spectrum division, moving average normalisation and airmass detrending which worsens this effect when included) are not applied to the models during retrievals. This mainly affect models that are almost constant with time: very broadened (hence fast rotating) spectra or planets with low semi amplitudes \footnote{For real planet, this effect is probably smaller because of the variability of the planet spectra, due to intrinsic variability and 3D geometric projections}. We are working on finding the most efficient way to include this additional model degradation in the Nested Sampling algorithm in order not to increase too much the numerical cost. Until then, our analysis is biased to lower molecular content for models with large rotation rate.

\subsection{Including clouds}
\label{ssec:clouds_synth}

Our next test was to study the influence of clouds in our data analysis which are a major limitation for atmospheric analysis (e.g., \citet{Kreidberg2014}). HRS has the potentiality to see above the clouds when they are deep enough \citep{Gandhi2020,Hood2020} and we first tried to recover synthetic planets with gray cloud deck at different pressure levels. When the cloud deck was below 0.1 bar, we recover the model roughly at the same amplitude than the non cloudy model. This is consistent with the fact that we expect to probe pressure levels around 10-1000 Pa through water absorption.  When we move the clouds higher up in the atmosphere, we typically lose one point of signal to noise detection per order of magnitude in pressure, until 0.1mbar where the SNR becomes lower than 2. However, some absorption lines are still theoretically observable \citep{Gandhi2020} and combining several observations might allow to push this limit upwards.

We have then ran a retrieval including clouds on a model with no clouds and obtained that, even if clear models exhibit higher likelihood, models with clouds are not excluded by our analysis: we obtain a degeneracy between water content and cloud coverage. This is not surprising as HRS is only sensitive to the variations of the atmospheric absorption, and not its absolute value. Additional constraints, such as LRS or fixed temperature can lift this degeneracy, as we will see in the application to real data on section \ref{sec:HD189}.

\begin{figure*}
    \centering
    \includegraphics[width=\linewidth]{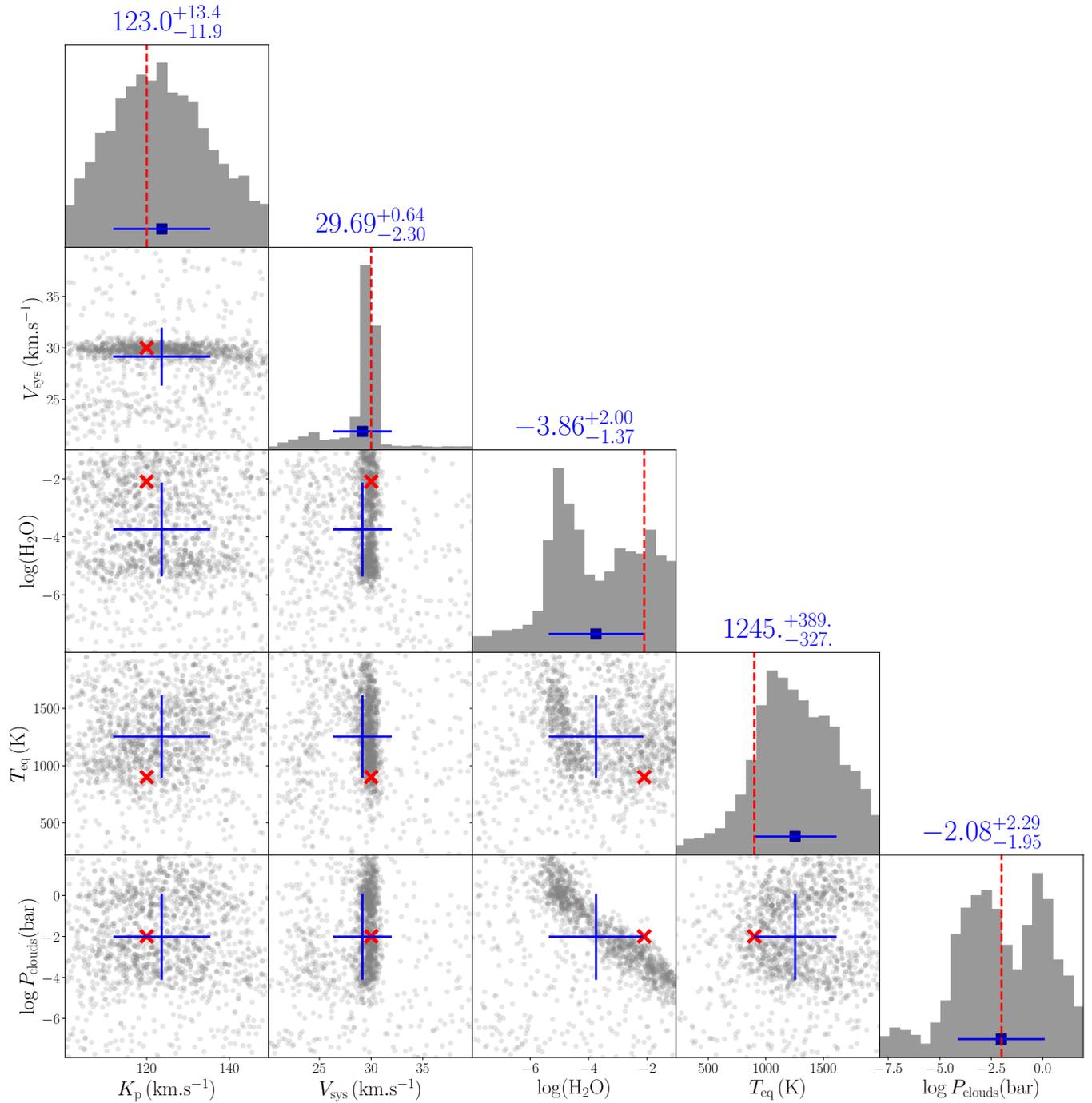}
    \caption{Corner plot showing the results of a \texttt{pymultinest} retrieval with a model containing only water with a mass mixing ratio of $10^{-2}$ and gray cloud deck at 10mbar. The different figure elements are the same as in Fig.~\ref{fig:nested_simple} with $P_\mathrm{clouds}$ being the recovered cloud top pressure.}
    \label{fig:marg_cloud2}
\end{figure*}

Finally, we have created a model with a gray cloud deck at 10mbar and applied our multinest algorithm which results are shown on Fig.\ref{fig:marg_cloud2}. Globally, the fit is poorer which is expected as the amplitude of planetary lines is reduced. We recover a very tight degeneracy between water and cloud top pressure spanning almost 5 orders of magnitude in water, showing that we lose our capability to obtain a precise water mixing ratio in that case without further constraints. This will be discussed again in section \ref{sec:HD189} where the change in the temperature profile allows to lift part of this degeneracy. This confirms our test with the non cloudy model: the use of HRS alone does not allow to lift the cloud-composition-temperature degeneracy, and additional information must be added. However, we note that our model does recover the injected model at the 1$\sigma$ level, which validates our pipeline for (simple) cloudy models.

\section{Application on real data}
\label{sec:HD189}

\subsection{Short description of the data}
In order to validate entirely our pipeline, we have applied it on real data with already published result for comparison. We have therefore used the two transits of HD~189733 b observed by SPIRou  as in \citet{Boucher2021} (hereafter B21). We shortly detail these data here, a larger discussion can be found in B21. The physical parameter  of the planet are referenced in Table \ref{tab:parameters}. 

Two transits of HD 189733 b were observed as part of the Spirou Legacy Survey (SLS, PI: J.-F. Donati). The first transit was observed on UT 2018 September 22 (hereafter Sep18), as part of SPIRou commissioning observations, and the second  on UT 2019 June 15 (hereafter Jun19). The first data set consists of 2.5 hr, divided into 36 exposures, where the first 21 are in transit and the remaining 15 are out-of-transit. The second data set consists of 50 exposures in total, where 24 are in transit, 12 before, and 14 after transit, for a total of $\sim$3.5 hr. The data were reduced using APERO version 0.6.132. In both observations, the airmass remains below 1.3 and even below 1.15 during the transit. The mean signal to noise ratio per order ranges from 50 in the telluric contaminated region and in the bluest part of the instrument to 250 in the center of the H band. Conditions were photometric for both transit sequences, with an average seeing of around 0$^{"}$.82 as estimated from the guiding images.

\subsection{Data analysis and retrieval}

We have therefore applied our pipeline on these two transits of HD 189733 b in order to retrieve atmospheric signatures and compare with B21. In order to be as consistent as possible with their methods, we have used the additional telluric correction of B21 (i.e. masking telluric lines deeper than 70\% from the continuum level, as described in Section~\ref{ssec:analysis}), and the detrending with airmass was only performed through PCA (see Section \ref{sec:Methods}). However, as presented in Section \ref{sec:Methods}, our pipeline has some intrinsic differences with B21, notably (i) we interpolate only the reference spectrum, and not individual ones, and (ii) the number of PCA components to remove is decided automatically by the pipeline. In both data sets, the number of PCA removed ranges from 1 in the bluest orders to 4/5 in the reddest orders.

\begin{figure}
    \centering
    \includegraphics[width=\linewidth]{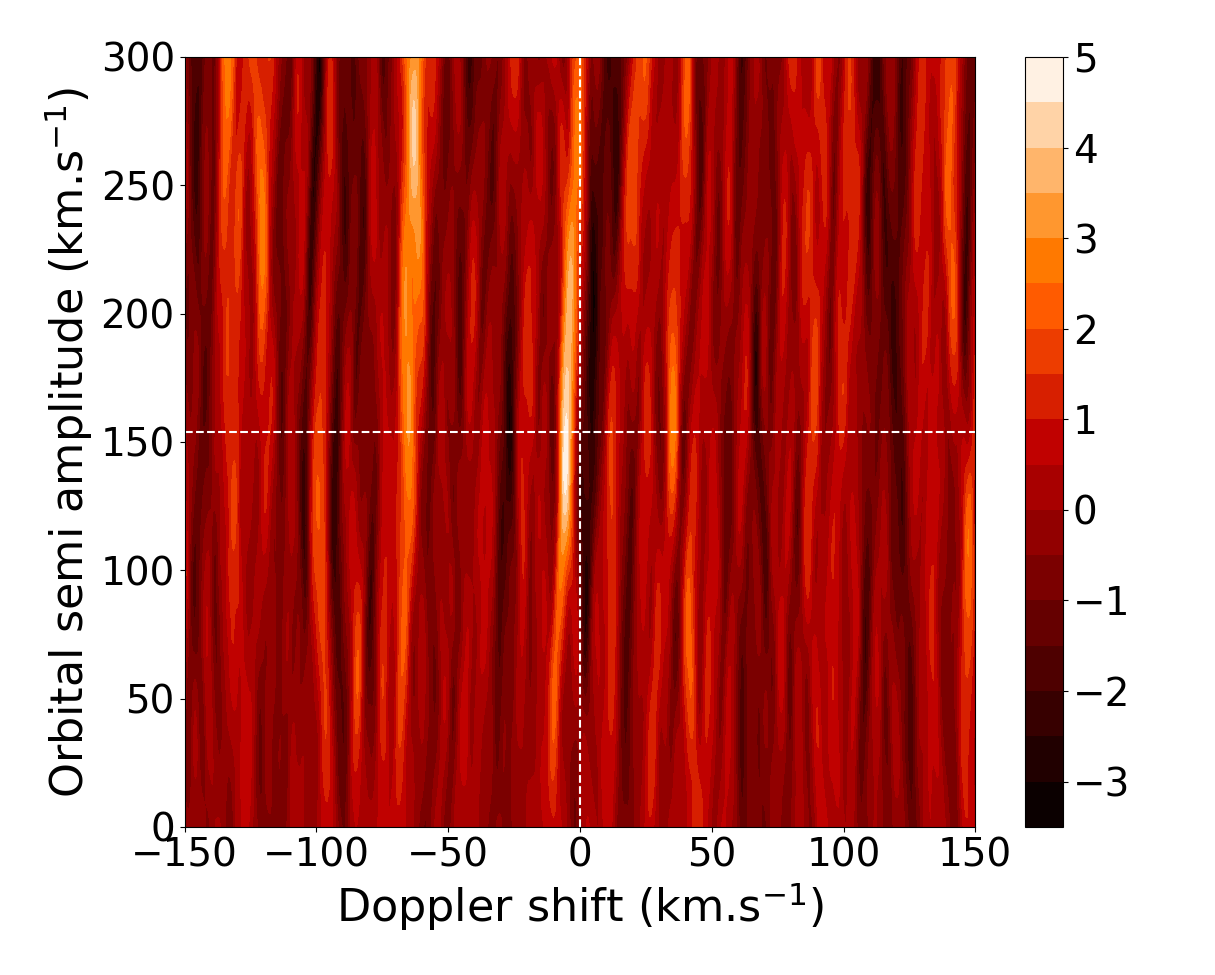}
    \centering
    \centering
    \includegraphics[width=\linewidth]{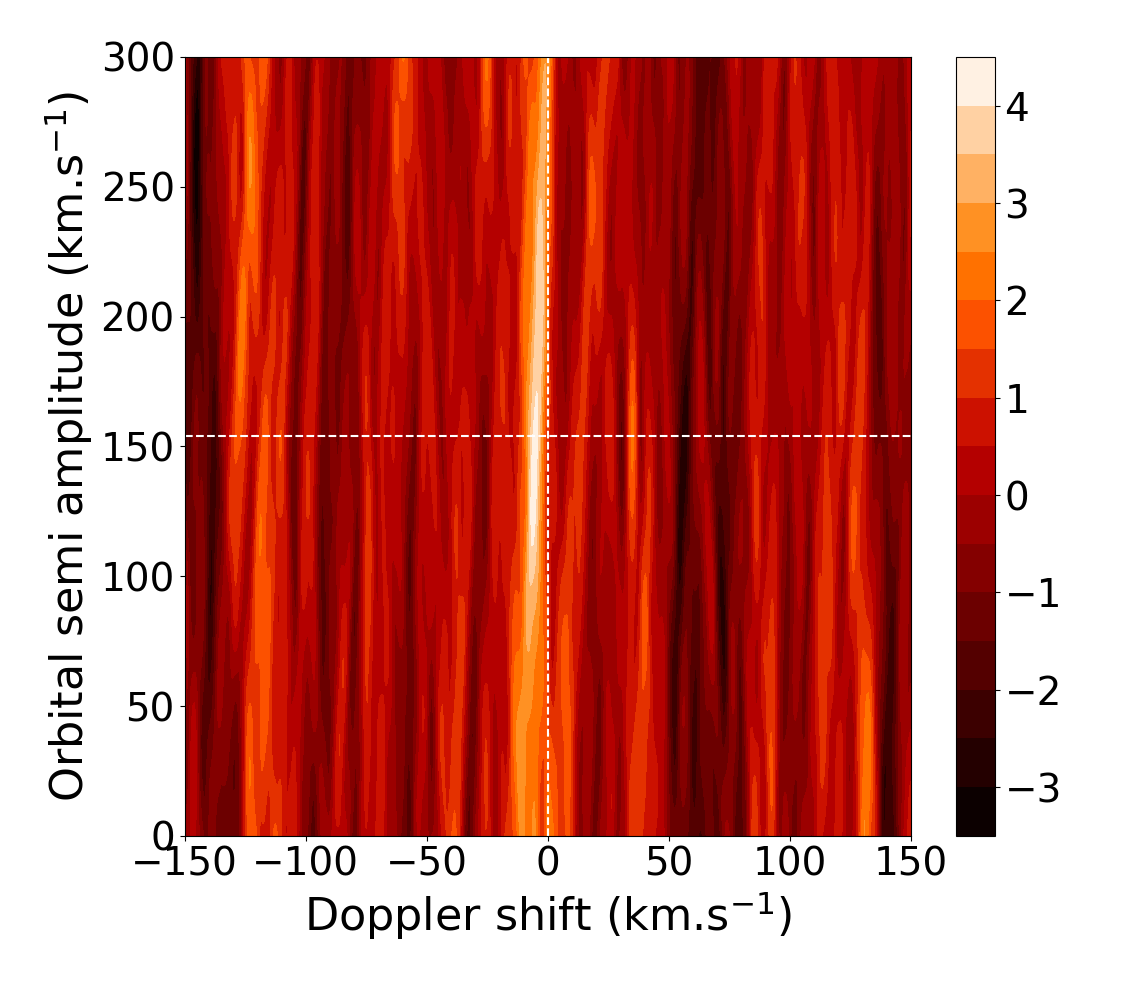}
    \centering
    \includegraphics[width=0.95\linewidth]{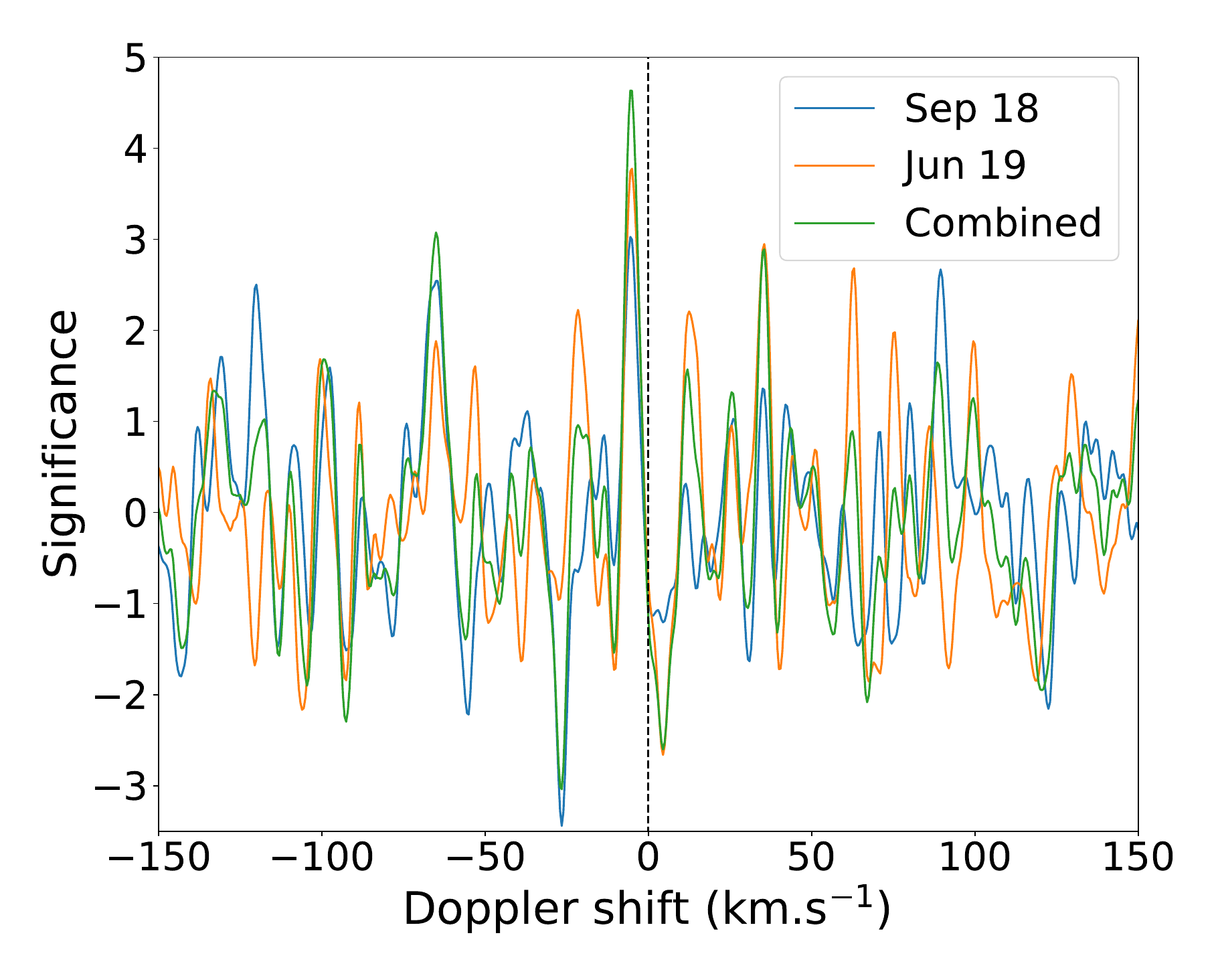}
    \caption{Top: cross-correlation map between the combined transits of HD 189733 b analyzed with PCA and the atmospheric model used in B21. The white dashed line show the theoretical position with no atmospheric Doppler shift. Middle: Same with the auto-encoder. Bottom: cross-correlation significance from the PCA-reduced data for individual transits and both transits and an orbital semi amplitude equal to the planetary semi amplitude (151.2 km.s$^{-1}$). The black dashed line is the 0 Doppler shift. } 
    \label{fig:ccf_boucher}
\end{figure}

In the template matching algorithm, we have used the exact same model than the best model of B21. The resulting cross-correlation map is shown in Fig. \ref{fig:ccf_boucher} to be compared with their Figure 5. The comparison is excellent: the maximum is recovered at the expected theoretical semi-amplitude (151 km.s$^{-1}$) and the recovered Doppler shift is comparable within B21 with less than 500m.s$^{-1}$ of difference. We obtain a slightly higher maximum of correlation (SNR of 4.6 compared to $4$ for B21) with a lower amplitude for a same non planetary peak obtained in both our works at $K_p \sim 270 $ km.s$^{-1}$ and $V_\mathrm{sys} \approx -75$ km.s$^{-1}$, showing that our pipeline corrects better for spurious signatures. We also applied our autoencoder on these data and found that the detection was slightly lower (SNR of 4.2) but exactly at the same position and the secondary spurious peak disappeared, confirming that it is not a physical signature. Additionally, the negative maximum of correlation next to the positive maximum, which is often recovered in studies using PCA disappears with the autoencoder. This is promising towards a more global use of this technique: it shows that coupling PCA and autoencoder can allow to disentangle between numerical and physical signatures which will be of real added values for planets with lower atmospheric detectability. 

\begin{figure*}
    \centering
    \includegraphics[width=0.95\linewidth]{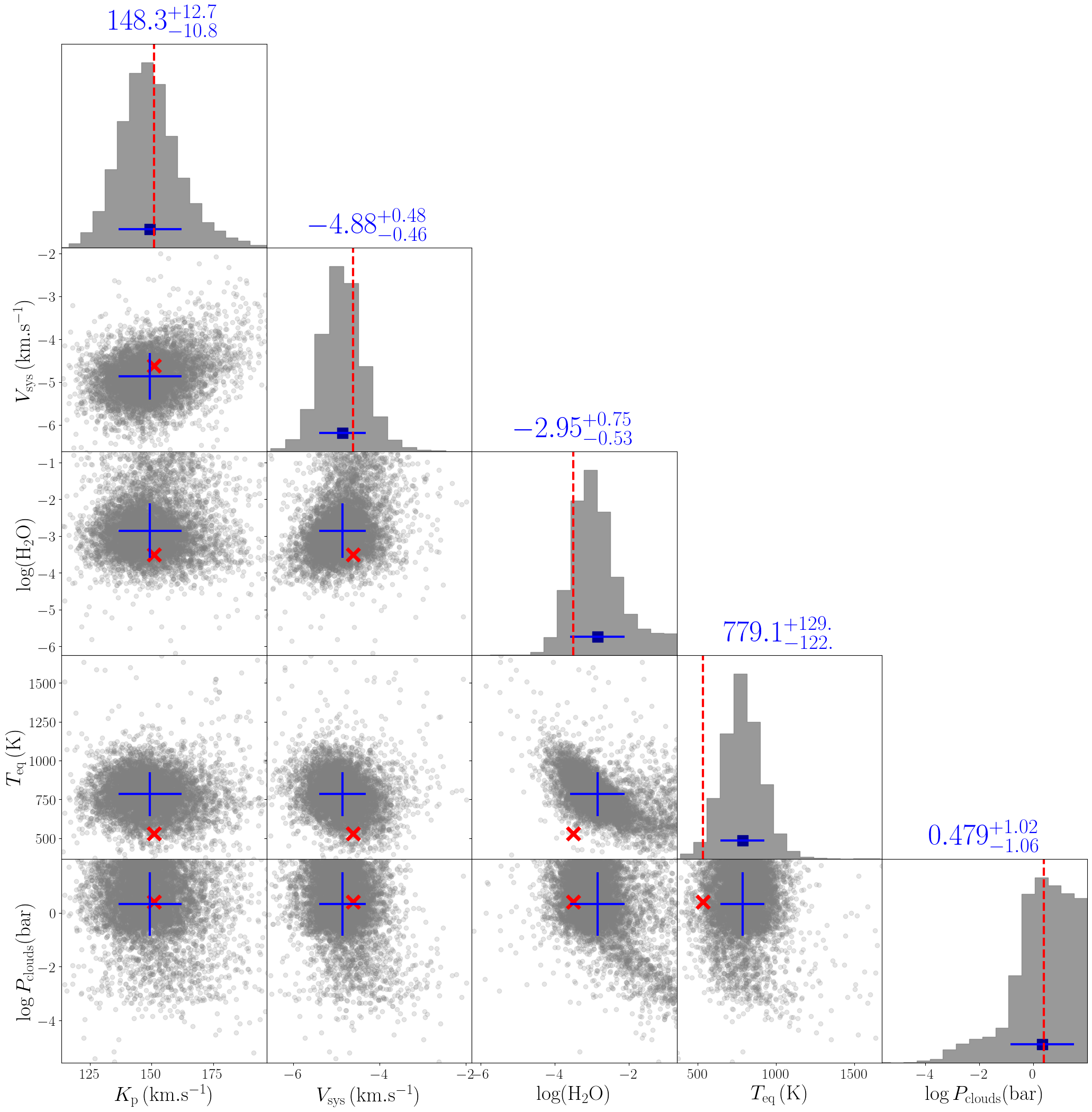}
    \caption{Posterior density for our nested sampling algorithms applied on the two transits of HD~189733 b with a temperature profile from \citet{Guillot2010}. The red crosses are the B21 recovered values. The water quantity is in mass mixing ratio, contrary to B21 in volume mixing ratio.}
    \label{fig:marg_boucher}
\end{figure*}

We have then used our nested sampling framework to retrieve parameters and compare with Fig. 11 of B21. We created models of HD~189733 b using petitRADTRANS with the water line list POKAZATEL from Exomol \citet{Polyansky2018,Tennyson2016} and a temperature profile from \citet{Guillot2010} as in B21. The resulting posterior distributions is shown in Fig.\ref{fig:marg_boucher} with the red crosses being the mean recovered values of B21. We see that we are perfectly consistent with their recovered parameters although we recover a higher temperature (which is more consistent with physical expectation of the temperature at the limbs of HD 189733 b (e.g., \citet{Drummond2018})), a higher water content and deeper cloud top pressure. Comparing to our test on synthetic data, it is striking how well the deep cloud top pressure is recovered. This surprising good retrieval led us to consider an isothermal profile as shown in Fig.\ref{fig:marg_boucher_clouds_iso}. As we expected, our algorithm then does not distinguish between a high cloud deck-high water content and deep cloud deck-low water content. We indeed see two gaussian distributions in the water posterior density, as in Fig. \ref{fig:marg_cloud2}. This shows how different variables are intricated, as we explore further in our companion paper, and how adding information on the temperature can lift the degeneracies in other parameters. As already mentioned, a combination with low resolution spectroscopy would also allow to solve this discrepancy: the observation of the slope of the continuum favours clear atmospheres \citep{sing2016,Barstow2020}.

In Appendix \ref{app:posterior}, we show two other posterior distributions when including rotation: one where the rotation speed is imposed as the expected tidally locked value for HD 189733 b (2.6 km.s$^{-1}$ at the equator) and one where the rotation speed is left as a free parameter. In both cases, we did not include clouds as they only increase complexity and are disfavoured, as mentioned above. For the first one, Fig. \ref{fig:marg_boucher_rotated}, we obtain similar results than in Section \ref{ssec:rotation}: we recover a lower water MMR and a higher temperature with larger error bars.  This globally confirms that the 1D, non rotated water MMR is a good estimate as this is coherent with our analysis on simulated data. We also recover a higher $K_p$ and the systemic velocity is changed by 500m.s$^{-1}$, being then perfectly in line with B21.  In the second one, Fig.\ref{fig:marg_boucher_rotatedfree}, the error bars are expectedly larger as rotation adds another degeneracy but we still recover consistent parameters and show that the data are consistent with tidally locked rotation. In summary, this application to real data confirms the validity of our methods globally. 

\section{Discussion}\label{sec:disc}

\subsection{Errors on K$_p$}
\label{ssec:disc-Kp}


In our simulations (and in general HRS planet observations), the absorption lines of the planet atmosphere are largely drowned in the noise, which limits the achievable precision of our recovered parameters. From our simulated data, e.g., Fig. \ref{fig:nested_simple}, we see that the 3 $\sigma$ error on the velocity is of the order of a third of a SPIRou pixel, and the 3 $\sigma$ error on $K_p$ leads to a shift of half a pixel at the beginning and end of the transit. These are extremely simplified cases as the injected and recovered model are very similar and it is therefore expected that, for real data, the error on \kp\ can be a factor of a few larger. It does however provide a good understanding of the precision of the method: being limited at the half pixel precision points towards the fact that we predominantly recover the center of the lines and not their shapes. 

Another interesting aspect is that, although we have performed a lot of simulations, we never recovered a mean \kp\ value that was lower than the injected \kp\ in synthetic data. In contrast, the broadening induced by the rotation can lead to a recovered mean \kp\ more than $20$\,\kms\ higher than the injected value. We attribute this effect to the data reduction process: the division by the median as well as the PCA/autoencoder aims at suppressing signals that are almost constant in time over the whole sequence, hence that have low \kp\ values. The PCA is also applied to the models, which reduce this effects, but not the first steps of the data analysis which have a lower but non zero impact on the model. This artificially enhances the recovered \kp\ and is coherent with the fact that this trend gets worse with higher rotation rates: the broadening of the lines make them more sensitive to the data analysis. The Doppler shift between the beginning and end of transit in our fiducial sequence is about 12 \kms: a line broadened by a rotation kernel of a few \kms\ will be more affected by the pipeline than non-broadened lines. Finally, we note that \vsys\ is well recovered by our model, confirming that blueshifts (or redshifts) in observed data will most likely be of physical origin (i.e. atmosphere dynamics such as winds). 

Globally, the fact that this technique performs better at high \kp \ was known and expected: a better separation between planetary, stellar and telluric lines during the course of the transit increases the level of detection. This will be a challenge for e.g., PLATO targets \citep{rauer2014} which will have no detectable Doppler variation between the star and planet during transit. Through our consortium, we have obtained part of the transit of HIP 41378f, an inflated Neptune mass planet on a 1.5 year orbit (see e.g. \citet{Alam2022}), that shift by less than a meter per second from the stellar line over the course of the transit. It will be a good test to optimize our method for planets with low \kp, or a proof of the absolute limitation of this method in the context of slowly displacing planets.

\subsection{Water-temperature degeneracy}
\label{ssec:degen}
As we expected, there is a degeneracy between composition and temperature. Although their physical effect on the atmosphere is different, both these parameters affect strongly the amplitude of the lines: temperature through a change in pressure scale height and composition through a change in opacity (and to a lesser extent to the pressure scale height as well). It is however important to notice that our analysis is not biased: although the maximum of likelihood does not necessarily correspond to the injected model, the injected parameters do lie in the ellipse of recovered parameters. A naive way to reduce the impact of this degeneracy is to provide informative priors but one obviously has to be careful about their physical motivation. Essentially, we need other diagnostics to lift the degeneracies.

Two main diagnostics come into mind: (i)~visible observations, where the strength of the absorption lines such as Ca+, Na or FeI is so much larger than it can provide tighter constraint on the rotation or temperature through the line shape, and (ii)~combination with low resolution spectroscopy (LRS). LRS can provide a reference value for the strength of the absorption line, hence a tighter constraint on the temperature-composition degeneracy. This in turn will provide a more appropriate estimate of dynamical mechanisms in the planet. 

\subsection{Degrading the model}
\label{ssec:disc-degrade}

As we discussed in section \ref{sec:intro}, we degrade the model because of the PCA treatment accordingly to the fast method developed by \citet{Gibson2022}. However, as mentioned in section \ref{ssec:rotation} and discussed in Appendix \ref{app:bias}, we do not degrade the model from the data analysis steps prior to PCA which leads to errors for broadened models. This is work in progress and is not discussed further here. 

Regarding the PCA degradation, we have made several tests to try to understand how this step was important. As can be seen on Fig. \ref{fig:degraded}, the degradation first leads to reduce the depth of several absoption lines. When we do not degrade the model, we therefore expect to recover lower mixing ratio or lower temperature to mimic this effect. This is exactly what we obtained, and when we tried to recover only the water content the error could reach a factor of 20 with a non degraded model. Degrading the model correctly is therefore of primordial importance. 

Essentially, without this step of model degradation, our recovered values were much further away to the real data and sometimes incompatible at the $3\sigma$ level. The correction provided by \citet{Gibson2022} allows to correct these effects, while only costing a small amount of calculation time. We are looking for such an efficient method to apply with the auto-encoder.

\subsection{Perspectives of exploration}

As the goal of this paper is mainly to present the pipeline, we have focused on a few examples but we could obviously not cover all the issues in the analysis of planetary atmospheres. Our companion paper, Debras et al. submitted, tackles the importance of biases and degeneracies in a few different situations. There are however many aspects that we did not include in these first ATMOSPHERIX papers.

Probably the most important is that we have not mentioned at all 3D effects, although they are known to impact the retrieval of atmospheric parameters \citep{Flowers2019,Caldas2020,Pluriel2020,Pluriel2022,Falco2022}. It would have been too large a task to explore these effects for a benchmark study, and we dedicate it to individual planet studies with the forthcoming works of the ATMOSPHERIX consortium. 

As we mention in the introduction, the way forward in our understanding of planetary atmospheres is the combination of low- and high-resolution spectroscopy. The method to combine these observations efficiently has been presented and discussed in \citet{Brogi2017}, and an application with SPIRou data has already been performed \citep{boucher2023}. We are therefore working on a similar benchmark paper with the combination of space and ground based data, notably in the goal of exploiting at best the JWST and Ariel capabilities.  

Finally, we chose to focus on infrared transmission spectroscopy with SPIRou but this pipeline could in theory be easily adapted to emission/reflection spectroscopy or data from another instrument/wavelength range. Emission spectroscopy has already been performed with SPIRou observations \citep{Pelletier2021} with similar methods and we have gathered and reduced visible data from MAROON-X with our pipeline, requiring marginal changes. Therefore, our pipeline is straightforwardly applicable to a much broader range of observations, which will be presented in the future.

\section{Conclusion}\label{sec:conclusion}

In this paper, we have presented our publicly-available pipeline to (i)~generate synthetic SPIRou transmission spectroscopy data, (ii)~reduce the data with state-of-the-art methods including PCA and the use of an auto-encoder and (iii)~analyse the data to recover the injected planetary signal either through cross-correlation or statistical exploration of the parameter space in a Bayesian framework. We have also included a fast way to include and retrieve (super-) rotation in planetary atmospheres.

By creating synthetic sequences, we demonstrated the validity of our pipeline and explored its limitations. We have first confirmed that the auto-encoder was a working, independent method to combine with PCA to recover planetary signals. We have shown that our method is unbiased for simple 1D models, but some issues remain in the retrieval for models with large rotation rates. We have explored the impact of clouds, showing that they can also bias the results and require additional constraints to be properly recovered.

Importantly, we have confirmed that degrading the model was needed to ensure a proper retrieval. We have implemented the \citet{Gibson2022} method for PCA and are still working on a fast method for the auto-encoder and for the non-PCA steps of the data analysis as well. When the model is not degraded, the retrieved value of the mixing ratio in our tests could be more than 1 dex away from the input value, and the temperature about $200$K wrong.

Finally, we have applied our pipeline on real SPIRou observations of HD~189733 b and obtained slightly better results than the literature for the detection and characterisation of the atmosphere. We recover water at a SNR greater than 4.5 with a volume mixing ratio in line with the literature and a temperature profile consistent with physical priors. We also show that we are consistent with a tidally locked rotation rate, an important result for hot Jupiters.

In conclusion, we have benchmarked our pipeline for atmospheric observations of exoplanets. A companion paper (Debras et al., submitted) tests its limitations for non isothermal and multi-species models. With the ongoing observations of the ATMOSPHERIX consortium, we have proven to be ready for the challenge of atmospheric detection and characterisation and will present results from SPIRou observations in a suite of papers to come.

\section*{Acknowledgements}

Based on observations obtained at the Canada-France-Hawaii Telescope (CFHT) which is operated from the summit of Maunakea by the National Research Council of Canada, the Institut National des Sciences de l'Univers of the Centre National de la Recherche Scientifique of France, and the University of Hawaii. The observations at the Canada-France-Hawaii Telescope were performed with care and respect from the summit of Maunakea which is a significant cultural and historic site. BK acknowledges funding from the European Research Council under the European Union’s Horizon 2020 research and innovation programme (grant agreement no. 865624, GPRV). FD thanks the CNRS/INSU Programme National de Planétologie (PNP) and Programme National de Physique Stellaire (PNPS) for funding support. This work was supported by the Action Spécifique Numérique of CNRS/INSU. This work was granted access to the HPC resources of CALMIP supercomputing center under the allocation 2021-P21021. JF.D, C.M, I.B. X.B, A.C., X.D.,G.H., F.K acknowledge funding from Agence Nationale pour la Recherche (ANR, project ANR-18-CE31-0019 SPlaSH). AM, BC, BB, SV acknowledge funding from Programme National de Planétologie (PNP) and the Scientific Council of the Paris Observatory. X.D. and A.C. acknoweldge funding by the French National Research Agency in the framework of the Investissements d’Avenir program (ANR-15-IDEX-02), through the funding of the “Origin of Life” project of the Université de Grenoble Alpes.JL acknowledges funding from the European Research Council (ERC) under the European Union’s Horizon 2020 research and innovation programme (grant agreement No. 679030/WHIPLASH), and from the french state: CNES, Programme National de Planétologie (PNP), the ANR (ANR-20-CE49-0009: SOUND). JFD and BK acknowledge funding from the European Research Council (ERC) under the H2020 research \& innovation programme (grant agreement \#740651 NewWorlds).  Part of the work has also been carried out in the frame of the National Centre for Competence in Research PlanetS supported by the Swiss National Science Foundation (SNSF). WP acknowledge financial support from the SNSF for project 200021\_200726. PT acknowledges supports by the European Research Council under Grant Agreement ATMO 757858. E.M. acknowledges funding from FAPEMIG under project number APQ-02493-22 and research productivity grant number 309829/2022-4 awarded by the CNPq, Brazil. Finally, we thank the anonymous referee for valuable comments and suggestions throughout the paper.

\section*{Data Availability}
The data are available upon request to the author and the code to analyze them is publicly available.



\bibliographystyle{mnras}
\bibliography{biblio} 




\appendix

\section{Convoluting winds}
\label{app:winds}
In order to derive simple equations for including the effects of planet rotation and winds in our retrieval, one has to remember that the observable quantity is the transit depth, i.e. the area of the projected planetary disk over the stellar surface, as a function of frequency. The elementary area of a curve $r(\theta)$ between $\theta$ and $\theta+\mathrm{d}\theta$ is the area of the elementary triangle, hence $1/2 r \times r \mathrm{d}\theta$. Assuming that the center of the planet is the center of our polar coordinate system, the projected area of the planet at a frequency $\nu$, defined as the curve $R_p(\nu, \theta)$ on the stellar disk is:

\begin{equation}
    \mathcal{A}(\nu) = \dfrac{1}{2}\int_{-\pi}^{\pi} R_p^2(\nu,\theta)\mathrm{d} \theta,
\end{equation}
where $-\pi$ corresponds to the evening limb, $-\pi/2$ is the South pole, $0$ the morning limb, and $\pi/2$ the North pole of the planet. We can define an effective radius, $R_\mathrm{eff}$, such that

\begin{equation}
        \mathcal{A}(\nu) = \pi R^2_\mathrm{eff} (\nu) = \dfrac{1}{2}\int_{-\pi}^{\pi} R_p^2(\nu,\theta) \cdot \mathrm{d} \theta.
\end{equation}

In this appendix, we make the assumption that the planet is properly characterized as a one-dimensional object whose radius only depends on the wavelength Doppler-shifted by rotation. However, we note that it would be very easy to decompose the planet into different regions within this framework (e.g. equatorial region and high latitudes), hence creating a pseudo 2D planet. Similarly, a latitude-dependent weight could be added on this integral to account for limb darkening effect, but we restrict to simple considerations here. 

Assuming that the planet rotates as a solid body with equatorial speed $\mathrm{v}_0$, the observed planetary radius at mid transit in the stellar frame at a frequency $\nu$ is simply the integral of the planetary radius Doppler-shifted by rotation: 

\begin{equation}
    R^2_\mathrm{eff} (\nu) = \dfrac{1}{2 \pi}\int_{-\pi}^{\pi} R_p^2 \left( \nu \cdot \left[ 1+\dfrac{\mathrm{v}_0 \cos \theta}{c} \right]  \right) \cdot \mathrm{d} \theta,
\end{equation}
where $c$ is the speed of light. Assuming North-South symmetry in the plane, we can remove the factor of $1/2$ and integrate from $0$ to $\pi$, i.e. morning limb to evening limb. For computational efficiency, we wish to express this integral as a convolution and we need to express it as a function of speed instead of wavelength in order to convolve it with the instrumental profile as in \cite{Brogi2016}. For a given frequency chunk of mean $\nu_0$ (e.g. spectrograph diffraction order), the corresponding velocity $\mathrm{v}$ can be simply expressed as

\begin{equation}
    \mathrm{v} = c(\dfrac{\nu}{\nu_0}-1),
\end{equation}

\noindent which leads to 

\begin{align}
    \nu (1+\dfrac{\mathrm{v}_0 \cos \theta}{c}) &= \nu_0 (1+\dfrac{\mathrm{v}+\mathrm{v}_0 \cos \theta}{c}+\dfrac{\mathrm{v} \mathrm{v}_0 \cos\theta}{c^2})  \\
    &\approx \nu_0 (1+\dfrac{\mathrm{v}+\mathrm{v}_0 \cos \theta}{c}).
    \label{eq:freq_vel}
\end{align}

\noindent
Within a SPIRou order, $v$ typically goes from $-5000$\,km.s$^{-1}$ to $5000$\,km.s$^{-1}$ and the equatorial rotation speed reaches no more than a few km.s$^{-1}$. Therefore, the approximation made in Eq.~\ref{eq:freq_vel} is verified and the error on the frequency is lower than the spectral resolution of SPIRou at the border of the order\footnote{Note in passing that such approximation is also implicitely performed in \citet{Brogi2016} although not explicitely written.}.
We can then write:

\begin{equation}
\label{eq:first_eff_rad}
    R^2_\mathrm{eff} (\mathrm{v}) = \dfrac{1}{\pi}\int_{0}^{\pi} R_p^2(\mathrm{v}+\mathrm{v}_0 \cos\theta)\mathrm{d} \theta.
\end{equation}

The observed squared planet radius $R^2_\mathrm{obs}$ is obtained by convolving $R^2_\mathrm{eff}$ with the instrumental profile. We therefore fall back onto the numerical result of \citet{Brogi2016} by writing

\begin{equation}
    R^2_\mathrm{obs} (\mathrm{v}) = \dfrac{1}{\sigma \sqrt{2 \pi}}\int_{-\infty}^{\infty} R^2_\mathrm{eff}(t) \mathrm{e}^{-\dfrac{(\mathrm{v}-t)^2}{2 \sigma^2}} dt.
\end{equation}

\noindent
The rotation kernel $K(\mathrm{v})$ at mid transit assuming no limb darkening is therefore: 

\begin{align}
    R^2_\mathrm{observed} (\mathrm{v}) &=
    \dfrac{1}{\pi}(R^2_p \ast K)(\mathrm{v})  \\
    &=\dfrac{1}{\pi} \left[R^2_p(t) \ast \dfrac{1}{\sigma \sqrt{2 \pi}}\int_{0}^{\pi} \mathrm{e}^{-\dfrac{(t+v_0 \cos\theta)^2}{2 \sigma^2}} d \theta \right](\mathrm{v}).
    \label{eq:conv1}
\end{align}

Although expression~\ref{eq:conv1} is quite simple, we can still improve its computational efficiency by writing it as two convolutions, which are extremely fast numerical operations. Denoting,
\begin{equation}
    x = - v_0 \cos \theta,
\end{equation}
we have
\begin{equation}
    \mathrm{d}\theta (x) = \dfrac{1}{v_0 \sqrt{1-(x/v_0)^2}}\mathrm{d}x,
\end{equation} 
which is properly defined for $\theta \in [0,\pi]$. We can therefore modify equation~\ref{eq:first_eff_rad} to obtain 
\begin{align}
     R^2_\mathrm{eff} (\mathrm{v}) &= \dfrac{1}{\mathrm{v}_0\pi}\int_{-\mathrm{v}_0}^{\mathrm{v}_0}  R_p^2 (\mathrm{v}-x) \dfrac{1}{\sqrt{1-(x/\mathrm{v}_0)^2}}\mathrm{d} x \\
     &= \dfrac{1}{\mathrm{v}_0\pi} [R_p^2(x) \ast \dfrac{1}{\sqrt{1-(x/\mathrm{v}_0)^2}}](\mathrm{v}).
    \label{eq:reff_finn}
\end{align}
where the convolution is defined between $-v_0$ and $v_0$ only. 



Note that, in practice, d$\theta$ is not defined in $x$\,=\,$\mathrm{v}_0$. However, since equation~\ref{eq:reff_finn} is a Riemann sum, only the following term of the integral cannot be computed:

\begin{align}
 \dfrac{1}{\mathrm{v}_0\pi}\int_{\mathrm{v}_\mathrm{final}}^{\mathrm{v}_0}  R_p^2 (\mathrm{v}-x) \dfrac{1}{\sqrt{1-(x/\mathrm{v}_0)^2}}\mathrm{d} x
\end{align}

\noindent
where $\mathrm{v}_\mathrm{final}$ is the edge of our array of integration. We can then assume that $R_p$ is constant in this interval $[\mathrm{v}_\mathrm{final}, \mathrm{v}_0]$ and simply calculate the theoretical value that is missing and correct it in our code for all frequencies. But actually, to a very good approximation, one can also assimilate this correction to a constant correction at all frequencies within an order as the variation of $R_p$ are negligible in front of the mean of $R_p$: the first order error is a constant shift of the recovered radius. As the input radius and rotated radius should have the same mean (easily verified theoretically), it is an excellent approximation to simply calculate this convolution and shift the result to its expected theoretical mean\footnote{Care must be taken in that case when including superrotation or other latitudinal dependent effect.}.

Finally,  the observed radius is simply the convolution by the instrumental profile as expressed above:  
\begin{align}
R^2_\mathrm{obs} (\mathrm{v}) &=  \left(R^2_\mathrm{eff} (t) \ast \dfrac{1}{\sigma \sqrt{2 \pi}}  \mathrm{e}^{-\dfrac{t^2}{2 \sigma^2}}\right)(\mathrm{v}) \\
& =\dfrac{1}{\mathrm{v}_0\pi} \left\{\left [R_p^2(x) \ast \dfrac{1}{\sqrt{1-(x/\mathrm{v}_0)^2}}\right](t) \ast \mathrm{e}^{-\dfrac{t^2}{2 \sigma^2}}\right\}(\mathrm{v}).
\label{eq:app:final_rad}
\end{align}

\noindent
We found that this expression was the fastest numerically speaking, and allows us to include a simple prescription for rotation in our parameter space exploration. Its comparison with \citet{Brogi2016} is excellent. Additionally, as mentioned previously, the inclusion of superrotation is extremely easy: one just has to separate the first convolution into a sum of convolutions onto smaller latitude ranges and adding a shift which is latitude dependent. In the same vein, one can separate the planet into two hemispheres of different composition/temperature profile straightforwardly.

\section{Biases with strong rotation and low semi-amplitude}
\label{app:bias}

In section \ref{ssec:rotation} and Fig. \ref{fig:marg_rotated}, we have shown that the retrieved water content was 3 $\sigma$ smaller than the injected model. We have tried different values for the rotation rate and water content and always obtained smaller retrieved water content when rotation was included, pointing towards a bias in the analysis. We have therefore performed many tests, listed below, and came to the conclusion that this issue arises from the data analisis, prior to PCA, which is not applied to the model, and is more important for signatures that are almost constant with time. Indeed: 
\begin{itemize}
\item We tested whether it was a signal to noise ratio issue: rotation lowers the amplitude of molecular lines and hence decreases the signal to noise ratio. We have therefore used models with much lower volume mixing ratio of water but no rotation which leads to comparable SNR with our high VMR rotated model. There was no biases in our low VMR models: the retrieval was worsen (higher error bars) but the mean retrieved values much closer to the injected values. 
\item We have then tested the influence of rotation and not surprisingly, the higher the rotation rate the larger the mismatch between the injected and retrieved planet properties. However, the rotation kernel is not the origin of the problem: when we simply include a gaussian broadening, we obtain similar biases.
\item We have also tested whether the error on the retrieved parameters could be due to an erroneous PCA correction. We have created models with a gravity divided by 4 (hence much larger atmospheric depth), which allowed to detect the molecular signatures without using PCA. The same trend was found (although to lesser extent): broadened models were biased. 
\item We noticed that the bias is increased for reduced data that containted a step of airmass detrending prior to PCA, compared to models where the airmass correction is performed by PCA.
\item We have created m models with a planetary radial velocity shift divided or multiplied by 4. When we increase (decrease) the semi amplitude, we reduce (increase) the bias. Typically, with \kp=400 km.s$^{-1}$ and a rotation speed of $3$ km.s$^{-1}$ at the equator, there is no bias anymore. Hence, this effect is more important for planetary models that don't shift much in wavelength with time.  
\item To confirm further that this is not just a SNR issue we have included a planet with a \kp multiplied by 4, increasing the SNR, but with 10 times less water in the atmosphere, decreasing the SNR. Cross correlation detects this planet with a similar significance than our reference model. In this case, we don't retrieve biases on the recovered parameters,  confirming that  the limiting factor is not the level of detection.
\item Finally, we have included the $a$ and $b$ value of equation \ref{eq:like} in the retrieval which were set to 1 in the rest of this paper. We observed that the retrieved $b$ was not equal to 1 but does not affect the retrieval at all (only the value of the likelihood). On the other hand, the retrieved $a$ was very close to $1$ for non broadened models and deviates largely from 1 (converging to 0.5) for broadened models, showing that there is a discrepancy between the injected and recovered model. Interestingly, when we fit the $a$ value, the bias is lowered but that is not a satisfactory solution to us and aim at providing the most reliable data analysis. 
\end{itemize}

In summary, the bias arises for broadened models, especially when the velocity shift of the planet’s atmosphere with respect to the star is lower than a few resolution elements over the course of the transit and disappears for synthetic planets with large semi amplitudes. It is increased when an airmass detrending step is included, but does not depend on PCA. Hence everything points toward the fact that the first phases of the data analysis, which are not applied consistently to degrade the models during retrieval are responsible for this bias.  We are working on efficient technics to include them. 

\section{Additional posterior densities for HD 189733 b}
\label{app:posterior}

\begin{figure*}
    \centering
    \includegraphics[width=0.95\linewidth]{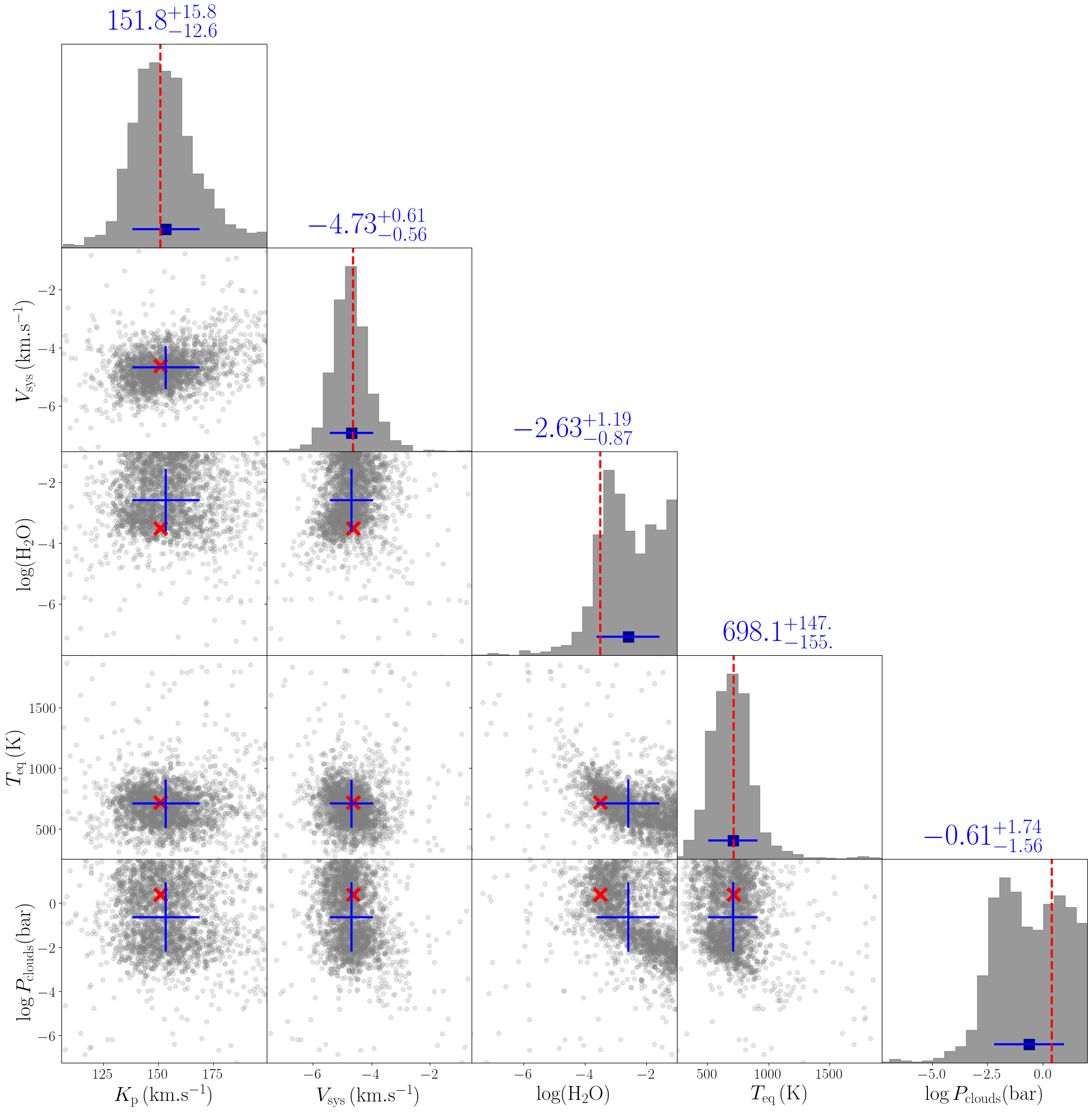}
    \caption{Same as Fig.\ref{fig:marg_boucher} but with an isothermal model. The temperature in red is the 1 bar temperature retrieved with the \citet{Guillot2010} profile in B21 }
    \label{fig:marg_boucher_clouds_iso}
\end{figure*}

\begin{figure*}
    \centering
    \includegraphics[width=0.95\linewidth]{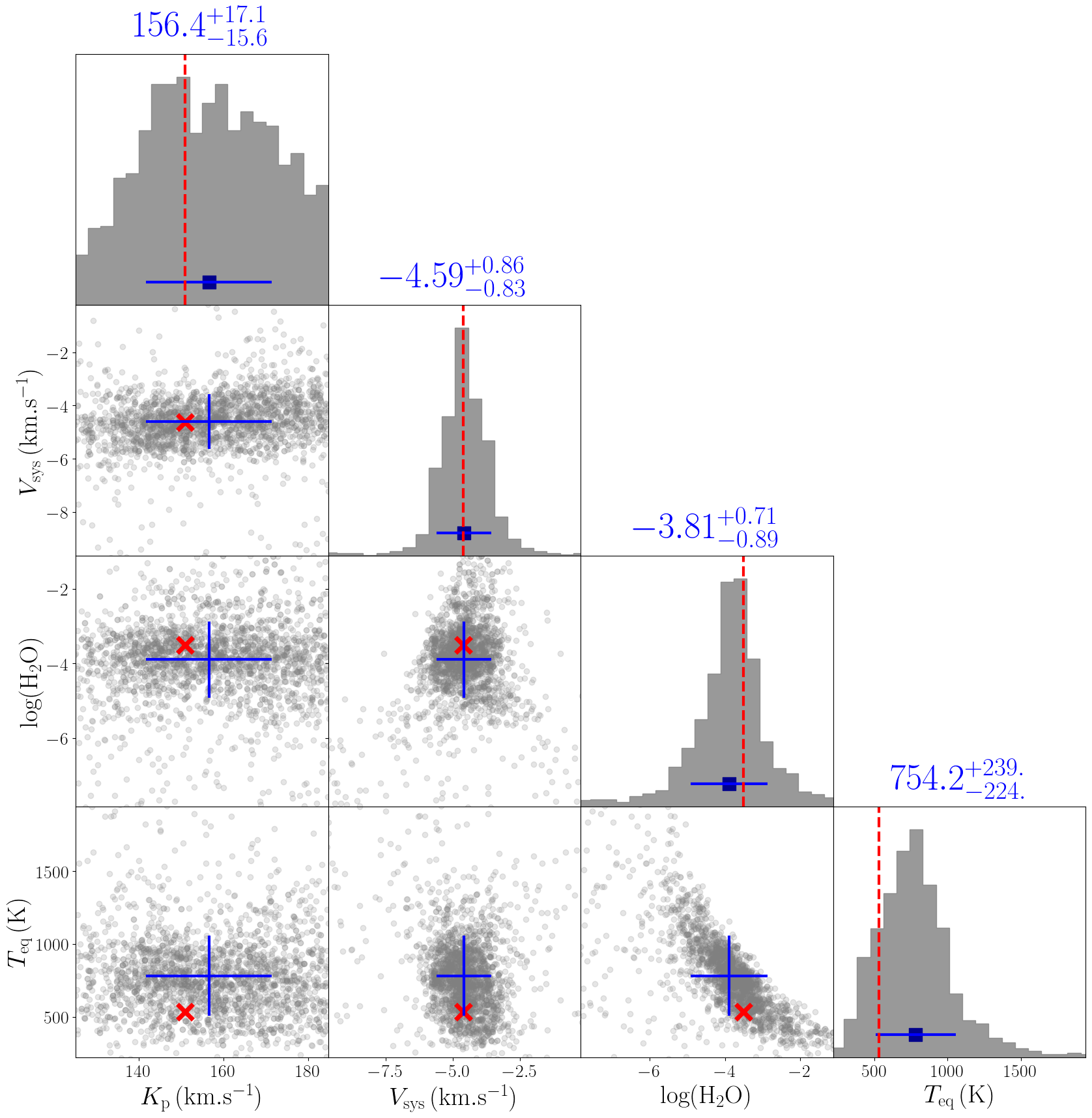}
    \caption{Same as Fig. \ref{fig:marg_boucher} but with the inclusion of a rotation in our model, with equatorial speed of $2.6$ km.s$^{-1}$.}
    \label{fig:marg_boucher_rotated}
\end{figure*}

\begin{figure*}
    \centering
    \includegraphics[width=0.95\linewidth]{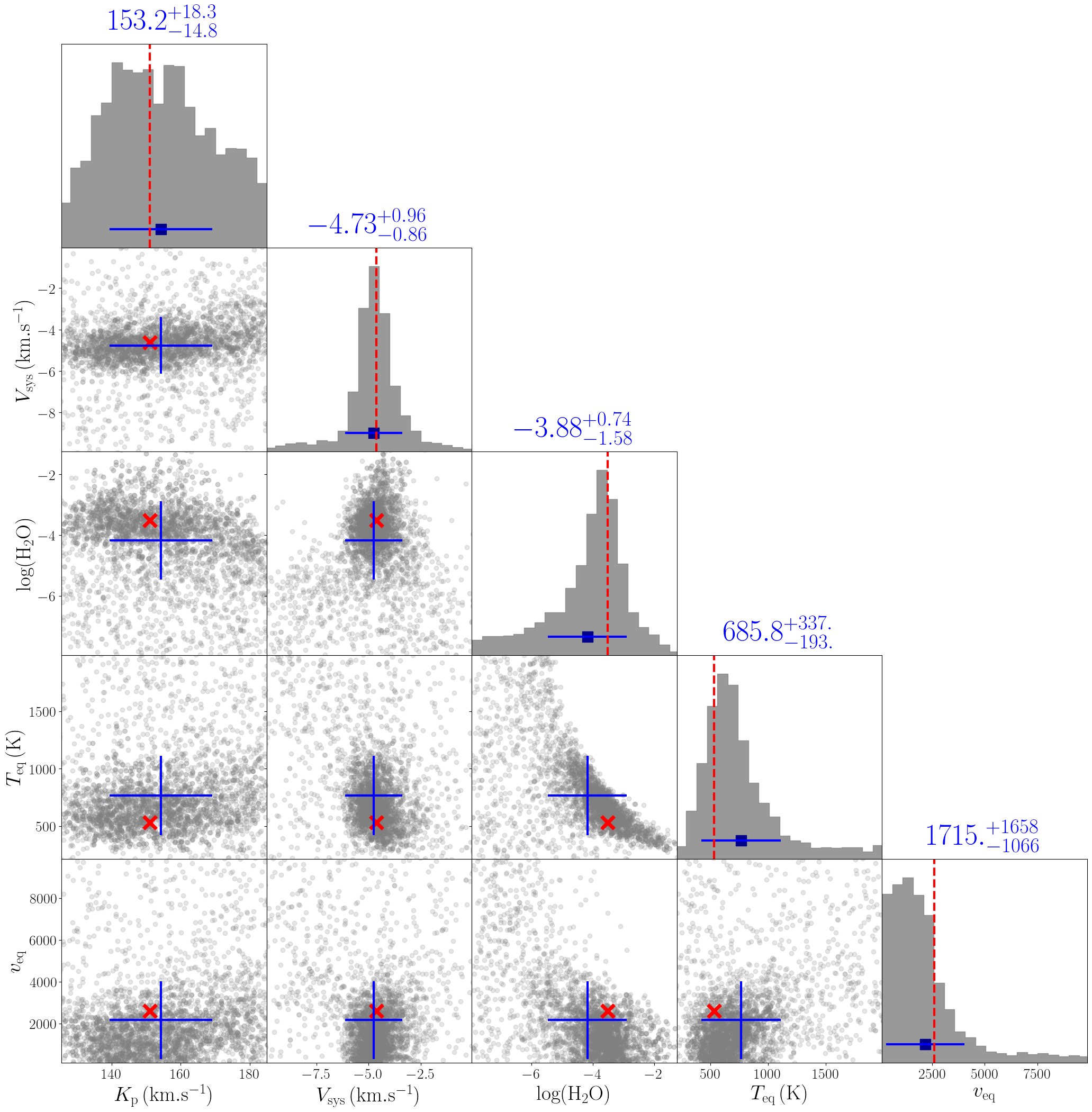}
    \caption{Same as Fig. \ref{fig:marg_boucher} but when the rotation is let as a free parameter and we display the rotation speed at the equator here. The tidally locked value is indicated by a red cross. }
    \label{fig:marg_boucher_rotatedfree}
\end{figure*}

\bsp	
\label{lastpage}
\end{document}